\documentclass[preprint,11pt]{elsarticle}
\usepackage{lineno,hyperref}
\usepackage{amsmath}
\usepackage{mathtools}
\usepackage{amssymb}
\usepackage{subfig}
\usepackage{soul}
\usepackage{graphicx}
\usepackage{hyperref}
\usepackage[letterpaper, margin=1.25in]{geometry}

\newcommand{\bl}{\begin{linenomath}}
\newcommand{\el}{\end{linenomath}}
\newcommand{\be}{\begin{equation}}
\newcommand{\ee}{\end{equation}}

\renewcommand{\tilde}{\widetilde}

\usepackage{xcolor}
\definecolor{backgreen}{rgb}{0.00, 0.169, 0.212}
\definecolor{textgray}{rgb}{0.514, 0.580, 0.589}

\usepackage{lineno,hyperref}
\usepackage{amsmath}
\modulolinenumbers[1]
\usepackage{verbatim}
\usepackage{amssymb}

\usepackage{xcolor}
\usepackage{mathtools}

\newcommand{\beq}{\begin{linenomath}\begin{equation}}
\newcommand{\eeq}{\end{equation}\end{linenomath}}
\newcommand{\bseq}{\begin{linenomath}\begin{equation*}}
\newcommand{\eseq}{\end{equation*}\end{linenomath}}
\newcommand{\bln}{\begin{linenomath}}
\newcommand{\eln}{\end{linenomath}}
\newcommand{\bal}{\begin{aligned}}
\newcommand{\eal}{\end{aligned}}
\newcommand{\bgat}{\begin{gathered}}
\newcommand{\egat}{\end{gathered}}

\newcommand{\E}{\mathbb{E}}

\newcommand{\MISE}{\mathrm{MISE}}
\newcommand{\ECDF}{\mathrm{ECDF}}
\newcommand{\IQR}{\mathrm{IQR}}
\newcommand{\bias}{B}
\DeclareMathOperator{\var}{Var}
\DeclareMathOperator*\argmin{\mathrm{arg \ min}}
\newcommand{\mcE}{\varepsilon}
\newcommand{\mcF}{\hat{f}}
\newcommand{\mcB}{\hat{B}}
\newcommand{\mcK}{\hat{K}}
\newcommand{\mcP}{\hat{P}}

\renewcommand{\bar}{\overline}

\biboptions{authoryear}

\journal{Advances in Water Resources}

\begin{document}


\begin{frontmatter}

\title{{
{Nonparametric, data-based kernel interpolation for particle-tracking simulations and kernel density estimation}}}



\author{David A. Benson\corref{mycorrespondingauthor}}
\address{Hydrologic Science and Engineering, Colorado School of Mines, Golden, CO 80401, USA}

\author{Diogo Bolster}
\address{Department of Civil and Environmental Engineering and Earth Sciences, University of Notre Dame, Notre Dame, IN 46556, USA}

\author{Stephen Pankavich and Michael J. Schmidt}
\address{Department of Applied Mathematics and Statistics, Colorado School of Mines, Golden, CO, 80401, USA}




\begin{abstract}
Traditional interpolation techniques for particle tracking include binning and convolutional formulas that use pre-determined (i.e., closed-form, parameteric) kernels.  In many instances, the particles are introduced as point sources in time and space, so the cloud of particles (either in space or time) is a discrete representation of the Green's function of an underlying PDE.  As such, each particle is a sample from the Green's function; therefore, each particle should be distributed according to the Green's function. In short, the kernel of a convolutional interpolation of the particle sample ``cloud'' should be a replica of the cloud itself.  This idea gives rise to an iterative method by which the form of the kernel may be discerned in the process of interpolating the Green's function.  When the Green's function is a density, this method is broadly applicable to interpolating a kernel density estimate based on random data drawn from a single distribution.  We formulate and construct the algorithm and demonstrate its ability to perform kernel density estimation of skewed and/or heavy-tailed data including breakthrough curves.
\end{abstract}


\end{frontmatter}

{\section{Introduction}}

In many applications, discrete samples of a continuous, and potentially complex, random process are generated as output, even though a continuous solution is desired.  Some examples are given by particle-tracking of passive solute transport \citep[e.g.,][]{Dani_2011,Pedretti_kernel,Siirila_2015,Morales_WRR_2018}, reactive particle transport \citep[e.g.,][]{dong_awr,Ding_WRR, Schmidt2017,Sole-Mari_2017,Sole-Mari_SPH,Sole2018,Benson_Poise,Perez-Poise,Engdahl_WRR, Engdahl_parallel}, and Monte Carlo simulation \citep[e.g.,][]{Tartakovsky_MCMC}.  A long history of statistical estimation has sought to best fit some continuous density function to a sequence of random samples, including maximum likelihood estimation \citep{Brockwell_Davis} and kernel density estimation \citep{Silverman1986}.  The former assumes a functional density form and estimates its parameters, while the latter fits a continuous function to discrete data, and tests of functional fits may be conducted later.  In hydrology (and many other sciences), the underlying processes being simulated may be sufficiently uncertain that a form for the density function cannot be assumed, and kernel density estimation is preferred.

{Given a true underlying pdf $f(x)$,} kernel density estimation is based on the convolution of discrete random data $\{x_1,x_2,...,x_n\}$ with some kernel function $K(x)$ producing the estimated pdf
\beq\label{eq:conv}
\bar f(x) = \frac{1}{\sum_{i=1}^n w_i} \sum_{j=1}^n \frac{w_j}{h_j}K\biggl( \frac{x-X_j}{h_j} \biggr),
\eeq
where $w_j$ are weights associated with data points $X_j$ (which could be prior ``concentrations'' that come from binning), $h_j$ are ``bandwidths'' associated with each kernel, and $K$ is some pre-determined, non-negative  function with the requirement $\int K(x)dx =1$ (i.e., $K$ is a pdf).  For random samples, the weights are equal constants that cancel from expression eq. \eqref{eq:conv}, {resulting in a factor of $1/n$}.  The common forms of $K$ are relatively simple (e.g., triangles or standard Gaussians) and yield estimates of $\bar f(x)$ with different properties such as regularity (i.e., number of derivatives) or compact support.  Kernels that are symmetric around $x=0$ are most commonly used (but certainly not always, see \cite{Asymmetric_book}), inasmuch as the eventual form of $\bar f(x)$, including skewness or heavy tails, are unknown {\em a priori}.  

It is well known that a pre-chosen kernel (such as a standard Gaussian) does not perform well if all of the bandwidths are chosen to be the same size.  Where data is more dense, the kernel bandwidths must be made smaller.  This has led to ``adaptive bandwidths'' that are adjusted based on the apparent or estimated density at the data points.  Higher estimated density values at data points are given smaller bandwidths.  But one may ask, should the functional form of the kernel also be adjusted based on the estimated density?  We suggest (and show in Appendix C) that the optimal kernel should be the same shape as the underlying true density, which is best estimated by the interpolated density.  But clearly, the estimated density changes if the kernel shape changes, therefore an iterative procedure is required.  We define this procedure in Section \ref{sec:procedure} after a brief review of kernel density estimation in Section \ref{sec:classic}.  A series of examples are given in Section \ref{sec:examples} and we conclude in Section \ref{sec:conclusion}.

\section{Classical Bandwidth Selection}\label{sec:classic}

Intuitively, one would like to choose a bandwidth as small as possible, because the convolution adds the variance of the kernels to the data itself.  On the other hand, as $h\rightarrow 0$, the kernels become delta functions and continuity of $\bar f(x)$ disappears.    Additionally, the choice of $h_i$ will depend strongly on both the eventual shape of $f(x)$ and the availability of random samples in any interval $[x,x+\Delta x]$. This has led to expressions that balance the bias and variance of the estimates.  A common place to start is to minimize the mean integrated squared error (MISE) between the estimated and unknown, real densities given by
\beq
\text{MISE}=\mathbb{E} \left [\int (f(x)-\bar f(x))^2 dx \right ] .
\eeq
Taking the expectation inside the integral and realizing that the mean squared error of an estimate is composed of squared bias and variance terms, {one finds}
$$\mathbb{E}[(f-\bar f)^2]=(\E [\bar f]-f)^2+\E [(\bar f -\E (\bar f))^2],$$
which gives a target {functional} for minimization.  Typically, a truncated Taylor series is used to {derive} asymptotic ($h\rightarrow 0, nh \rightarrow \infty$) expressions for the bias and variance that depend on the properties of the kernel and underlying density. {This process (Appendix A) results in approximations for the bias}
\beq\label{eq:bias}
B(x) = \text{bias}[\bar f(x)]=\E [\bar f(x)]-f(x)=\frac{h^2}{2}f^{\prime\prime}(x) \mu_2 (K) + \mathcal{O}(h^3),
\eeq
and variance
\beq\label{eq:var}
\var[\bar f(x)] = \E [(\bar f -\E (\bar f))^2]=\frac{1}{nh}f(x)\int K^2(x) dx + \mathcal{O}\left ((nh)^{-2} \right ).
\eeq
All other things held equal, letting $h \rightarrow 0$ minimizes bias, but variance grows without bound (i.e. accuracy increases but smoothness decreases), while letting $h$ grow large decreases the variance of estimates, but accuracy is sacrificed. Minimizing the sum gives a value for the optimal global bandwidth
\beq \label{eq:h_0}
h_{0}=\left( \frac{d\int K^2(x) \mathrm{d}x}{n (\mu_2 (K))^2 \int (f^{\prime\prime}(x))^2 \mathrm{d}x } \right) ^{1/(d+4)}
\eeq
where $d$ is the number of dimensions of the random variable ($d=1$ herein). {Notice that a finite second moment $\mu_2(K)$ is necessary to use this method in the estimation of the optimal bandwidth; we remove that requirement herein.} Without any information at all, it is common to assume Gaussian $f(x)$ and Gaussian kernels, in which case a constant bandwidth is used with size 
\beq\label{eq:Taylor}
h_0 \approx 1.06n^{-1/5}\hat \sigma,
\eeq
where $\hat \sigma$ is the sample variance. {Greater} data density means smaller bandwidth (until as $n\rightarrow \infty$, $h_0\rightarrow 0$).  Because this estimation of finite $h_0$ is based, in part, on an assumption of $h_0\rightarrow 0$, we might expect significant error in this estimate.  Indeed, an exact value of $h_0$ can instead be derived using the Fourier transform (Appendix B), and we show that the result in Eq. \eqref{eq:Taylor} can be significantly erroneous.

Furthermore, it is largely recognized \citep[e.g.,][]{Silverman1986} that the {\em local} data density is a better indicator of bandwidths that should be uniquely defined at each data point.  In regions where data density is smaller, the bandwidth should be greater.  There are several methods used to estimate local data density.  For example, \citet{Silverman1986} shows that for large $n$, the local data density can be approximated by the value of the estimated pdf, so that an adaptive bandwidth can be estimated by 
\beq\label{eq:adapt}
h_i = h_0 \lambda_i = h_0 \biggl( \frac{\tilde f(x_i)}{G} \biggr)^{-\xi},
\eeq
\noindent where the tilde indicates some intermediate estimate of the density, and the normalization factor $G$ is the geometric mean of estimated density values, namely
\beq\label{eq:G}
G=\exp\biggl( \frac1n \sum_{i=1}^n \ln \tilde f(x_i) \biggr).
\eeq
The exponent $0\le \xi \le1$ is an empirical weighting factor shown to be 0.5 under ideal conditions \citep{Abramson}.


In a novel way, \citet{Pedretti_kernel} investigated the use of the adaptive kernel methods (eqs. \eqref{eq:h_0}, \eqref{eq:adapt}, and \eqref{eq:G}) for interpolating breakthrough curves (BTCs) for simulated push-pull, single-well tests with trapping in relatively immobile (low-velocity) zones.  These BTCs are noteworthy for {their} thin early tails and fat late tails, or rapid (steep) early breakthrough and delayed, power-law decline of concentration.  \citet{Pedretti_kernel} applied their density estimation {method} on binned histograms, so that several points are different in several respects that we will outline below.  But importantly, those authors found that adjusting the bandwidth based only on particle density tended to overly broaden the early BTCs in order to more properly represent the late tail.  \citet{Pedretti_kernel} then imposed a restriction on broadening the kernel bandwidth based on whether particles (concentrations) in equation \eqref{eq:conv} occurred early or late in the BTC. This, of course, means that the user must decide how the kernel must be adjusted.  But this is simply a side effect of choosing, {\em a priori}, a non-physical and symmetric kernel.  If each particle were treated as a single realization of the Green's function, then its highly skewed kernel would transfer little mass to earlier portions of the BTC, and no adjustment may be needed.
 
\section{Iterative Algorithm}\label{sec:procedure}
We show (Appendix C) that to minimize the MISE, the kernel of each particle for simulations of the Green's function should be a scaled version of the Green's function itself.  Of course the shape of the density is not known {\em a priori}. Therefore, the algorithm discovers the kernel shape and size recursively according to the {following} steps:
\begin{enumerate}
\item\label{item:h0} Build a candidate $\tilde f_0(x)$ using constant bandwidth $h_0$ and standard Gaussian kernel $K(x)=(2\pi)^{-1/2}\exp(-x^2/2)$.
\item Use $\tilde f_0(x)$ to interpolate values at data points $\tilde f(x_i)$.
\item Use the values $\tilde f(x_i)$ in eq. \eqref{eq:adapt} to estimate adaptive bandwidths for the Gaussian kernel and re-estimate $\tilde f_1(x)$.  This would end classical estimation. Set counter $\ell=1$.
\item\label{item:re} Use $\tilde f_\ell(x)$ as the new kernel $K_\ell(x)=\tilde f_\ell(x)$. 
\item\label{item:std} Adjust kernel $K_\ell$ to have zero mean and unit ``width''.
\item Use $\tilde f_\ell(x)$ to interpolate values at data points $\tilde f(x_i)$.
\item\label{item:lambdas} Use the values $\tilde f(x_i)$ in eq. \eqref{eq:adapt} to estimate adaptive bandwidths for the new kernel $K_\ell(x)$.
\item Use new kernel function and bandwidths to estimate $\tilde f_{\ell+1} (x)$.
\item Return to step \ref{item:re} until desired closure between $\tilde f_{\ell+1}(x)$ and $\tilde f_{\ell}(x)$. Upon closure,  $\tilde f_{\ell+1}(x)$ is the best estimate of $\bar f(x)$.
\end{enumerate}

The potentially tricky parts of the algorithm are associated with steps \ref{item:h0},  \ref{item:re},  \ref{item:std}, and \ref{item:lambdas}.  For step \ref{item:h0}, the distributional qualities of the data are unknown, so we use a Fourier transform algorithm to estimate the data density function (see Eq. \eqref{eq:epsdef} in the Appendix).  By assuming a Gaussian kernel, the initial $h_0$ can be {easily} estimated.
For step \ref{item:re}, it is important to use a numerical domain for $x$ that is wider than the data values, so that the kernel may extrapolate sufficiently before the smallest data point and after the largest. Furthermore, for widely-spaced and sparse data, the density of calculated points in $x$ must also be chosen to provide sufficient resolution.   For step \ref{item:std}, it is not always clear that the mean and standard deviation exist or are the proper scaling metrics for the iterated kernel.  A simple example is a stable density, which may have diverging moments, and also rescale differently from, say, a Gaussian density. Here, we suggest using the interquartile range of the data and the kernel for a reasonably close and reliable estimate of the scale of many densities.  For step \ref{item:lambdas}, we are now using a kernel that is thought to resemble the underlying density, so using $K\rightarrow f$ or $\bar f$ in \eqref{eq:bias}, \eqref{eq:var}, and \eqref{eq:h_0} will give different values of $h_0$ etc.  More on these points is provided below.

In order to find a ``standard'' kernel from the previous iteration's density estimate $\bar f$, the kernel must have zero mean (so that the subsequent addition of kernels has the same mean as the data).
The width of the kernel should be standardized, such as normalizing by a scale factor equal to the standard deviation of the data or central second moment of $\tilde f$.  However, many densities have diverging second moments, so a robust method must be found for situations in which the underlying density is unknown.  A quick survey of the interquartile rage (IQR) and the scale factor of many densities shows reasonably similar relationships.  For finite-variance distributions we find, for example, the Gaussian has $\sigma \approx \IQR/1.35$; the exponential $\sigma \approx \IQR/1.1$; the Laplace $\sigma \approx \IQR/0.98$. Infinite variance distributions with closed-form distribution functions (characterized by scale parameter $\sigma$) include the symmetric Cauchy with $\sigma \approx \IQR/2$ and the maximally-skewed, 1/2-stable Levy density with $\sigma=\IQR/9$.  Using MATLAB's routine for calculating the CDF of a stable law, we find that, for a maximally-skewed 1.5-stable, $\sigma \approx \IQR /2.13$.  With the exception of the Levy density, it is a reasonable approximation to say that the ``width'' of the density function {may} be standardized using $\sigma \approx \IQR/1.5$.  Therefore, in the following, to arrive at a ``standard'' density from the data-kernel, we numerically integrate the intermediate density $\tilde f _{\ell-1}$ to find $\IQR=x_{0.75}-x_{0.25}$, where 
\bl
\begin{equation*}
x_z = \min \left \{x_i \  \biggr | \ \Delta x \sum_{\ell = 1}^n \tilde f_{\ell-1}(x_i)  \geq z \right \}
\end{equation*}
\el
and simply shift and rescale the experimental density by its first moment $m_1$ and a generic multiple of the IQR so that 
\bl
\begin{equation*}
K_\ell(x) = \frac{1}{(\IQR/1.5)} \tilde f_{\ell-1} \left (\frac{x- m_1}{(\IQR/1.5)} \right ).
\end{equation*}
\el

As noted before, the kernel is allowed to change after each iteration, and this kernel is checked against the previous iteration's kernel.  Iteration is {terminated} when the kernel converges and the difference {in successive approximations} is {sufficiently} small.  Here, we chose to {discontinue the algorithm} 
when the $L^2$ difference between {successive} iterations is less then $10^{-8}$, where 
\bl
\begin{equation*}
L^2 = \sqrt{ \Delta x \sum_{i=1}^n \left | \tilde f_\ell(x)-\tilde f_{\ell-1}(x) \right|^2}.
\end{equation*}
\el

We only use those points on the regularly-spaced grid with kernel values greater than $10^{-10}$ to avoid small numerical negatives that may arise depending on the interpolation scheme (i.e., splines can return small negatives).  If the $L^2$ difference is found to {\em increase} between iterations, this indicates too large a global bandwidth $h_0$, so the value of $h_0$ from the previous iteration is decreased by a factor of 0.8 and iteration resumes.

\section{Examples}\label{sec:examples}
We investigate the {iterative} algorithm versus classical (assumed Gaussian kernel) methods for {four} types of data: 1) symmetric and thin-tailed; 2) maximally-skewed and exponentially-tailed, 3) Symmetric and heavy, power-law tailed; and, 4) maximally skewed and heavy, power-law-tailed.  The last is chosen because BTC data are strictly positive and often observed to fall off like $x^{-1-\alpha}$, where $\alpha$ is on the order of 0.5.  We also investigate how well the estimators perform over a large realization of random samples and a range of population sizes $n=\{100,1000,10000\}$, inasmuch as large particle numbers (and random arrival times) are typically used.  In each case we use estimates of the MISE to measure bias and variance of the estimated density versus a known density on a regular grid in $x$.  A numerical estimate of the MISE is given by an ensemble mean of the $L^2$ norm, namely
\bl
\begin{equation*}
\bar \MISE = \frac{\Delta x}{M} \sum_{m=1}^M \sum_{i=1}^n |\bar f_m(x_i)-f(x_i)|^2,
\end{equation*}
\el
for a set of $M$ realizations of data with an underlying density $f$ and the corresponding estimates of the density $\bar f_m$ on a common grid of estimation points $x_i$ with spacing $\Delta x$.  For each of the examples, we generate $M= 100$ independent realizations of data from {known} distributions in order to estimate the densities and resulting MISE (Table \ref{tab:rmse}).

\begin{table}[t]
    \centering

    \begin{tabular}{| c || c | c | c | c |}
    \hline

    \hline
       \textbf{$Kernel$}   & \textbf{Normal} & \textbf{Exponential} & \textbf{Cauchy} & \textbf{1.5-Stable} \\
    \hline

         Gauss w/$h_0$    & 0.0013  &    0.157&     0.0279 &    0.0197   \\
         Gauss w/$h_i$     & 0.0012  &     0.135 &    0.0243 &  0.0118 \\
         Gauss iter.            & 0.0016 &      0.127 &    0.0231 & 0.0100 \\
         $\bar f$ iter.          & 0.0014 &      0.073 &    0.0150& 0.0020\\
    \hline

    \hline
    \end{tabular}
        \caption{Computed ensemble MISE for various kernel estimates from 100 realizations of 1,000 random variables.  The first row is for uniform initial bandwidth $h_0$.  The second is for single-pass application of adaptive bandwidth $h(x_i)$.  The third is iterated Gauss kernel until closure.  The fourth is data-based kernel iterated to closure.}
    \label{tab:rmse}
\end{table}

\subsection{Gaussian data}

\begin{figure}
 \centering
 \includegraphics[width=5.5in]{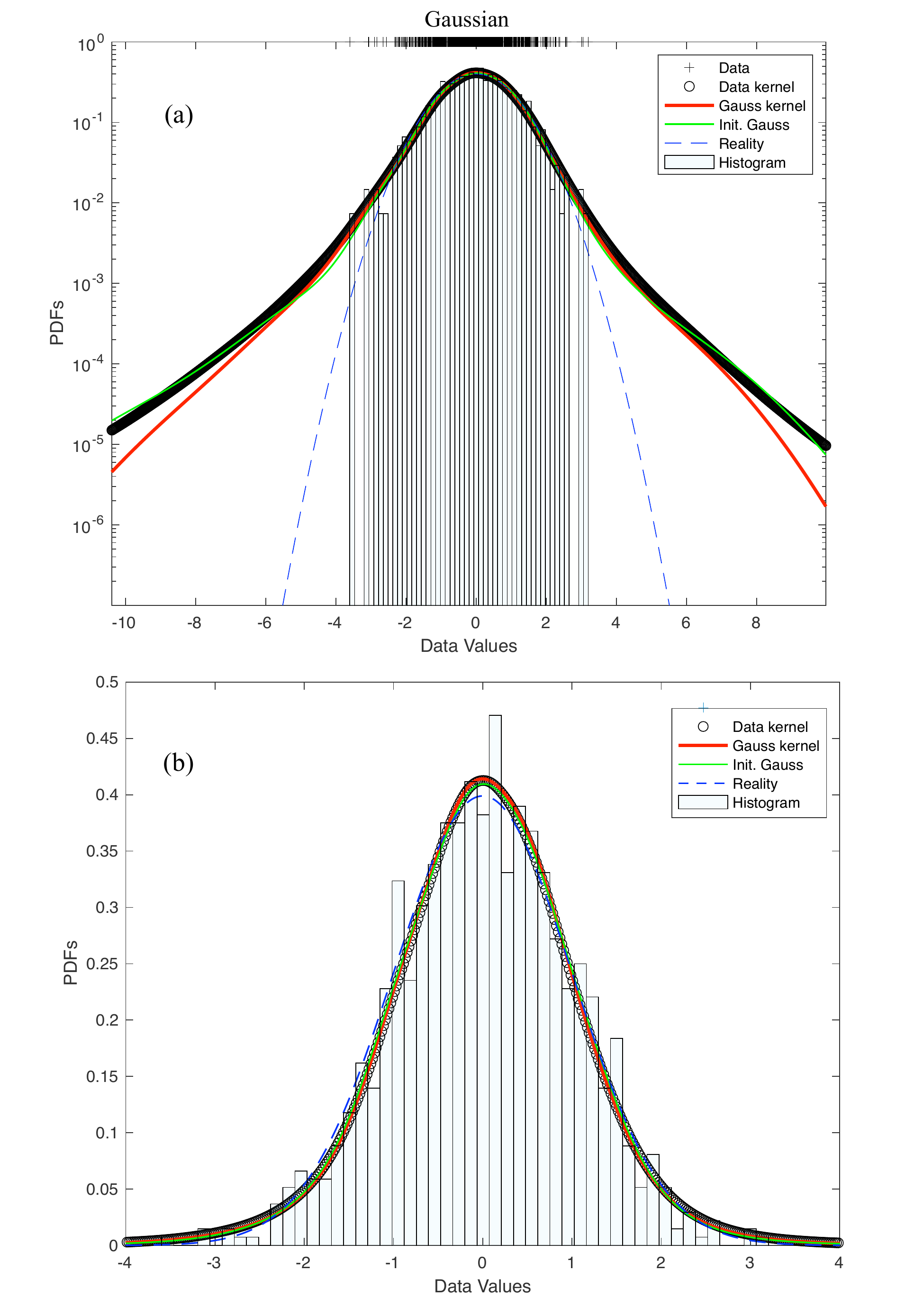}
 \caption{a) Semilog and b) linear plots of iteratively estimated densities for a single realization of 1,000 Gaussian data points using data-based kernel (black symbols) and Gaussian-based kernels (red curves).  Also shown are the single-pass Gaussian kernel estimate (green curves) and 20-bin histograms.  Blue dashed line is underlying ``true'' density function.}
 \label{fig:Gauss_1000}
 \end{figure}

We start with Gaussian random variables, in which the data kernel should (nearly) converge to an {\em a priori} Gaussian kernel, because the underlying data that builds the data-kernel is Gaussian.  Indeed, for a large number of data {points} (1,000), the iterated KDE for the data-based and Gaussian-based kernel are nearly identical, even in the extreme tails (Fig. \ref{fig:Gauss_1000}).  This example shows the robust nature of the estimation, inasmuch as the Gaussian kernel uses the exact width of the kernel $\sigma=1$, while the iterated kernel uses a general width estimate of $\IQR/1.5$.  In actuality, the width of a Gaussian is $\sigma = \IQR/1.34$.  It is worth noting that closure to the final kernel usually takes between 5 and 7 iterations. Furthermore, because the value of $h_0$ is not set exactly (which would require identifying the data as Gaussian before interpolating), the iterated kernels have similar magnitudes of MISE as single-pass adaptive Gaussian kernels and convolution with a single value of $h_0$ (Table \ref{tab:rmse}).

\subsection{Exponential data}\label{sec:exp}
\begin{figure}
 \centering
 \includegraphics[width=5.5in]{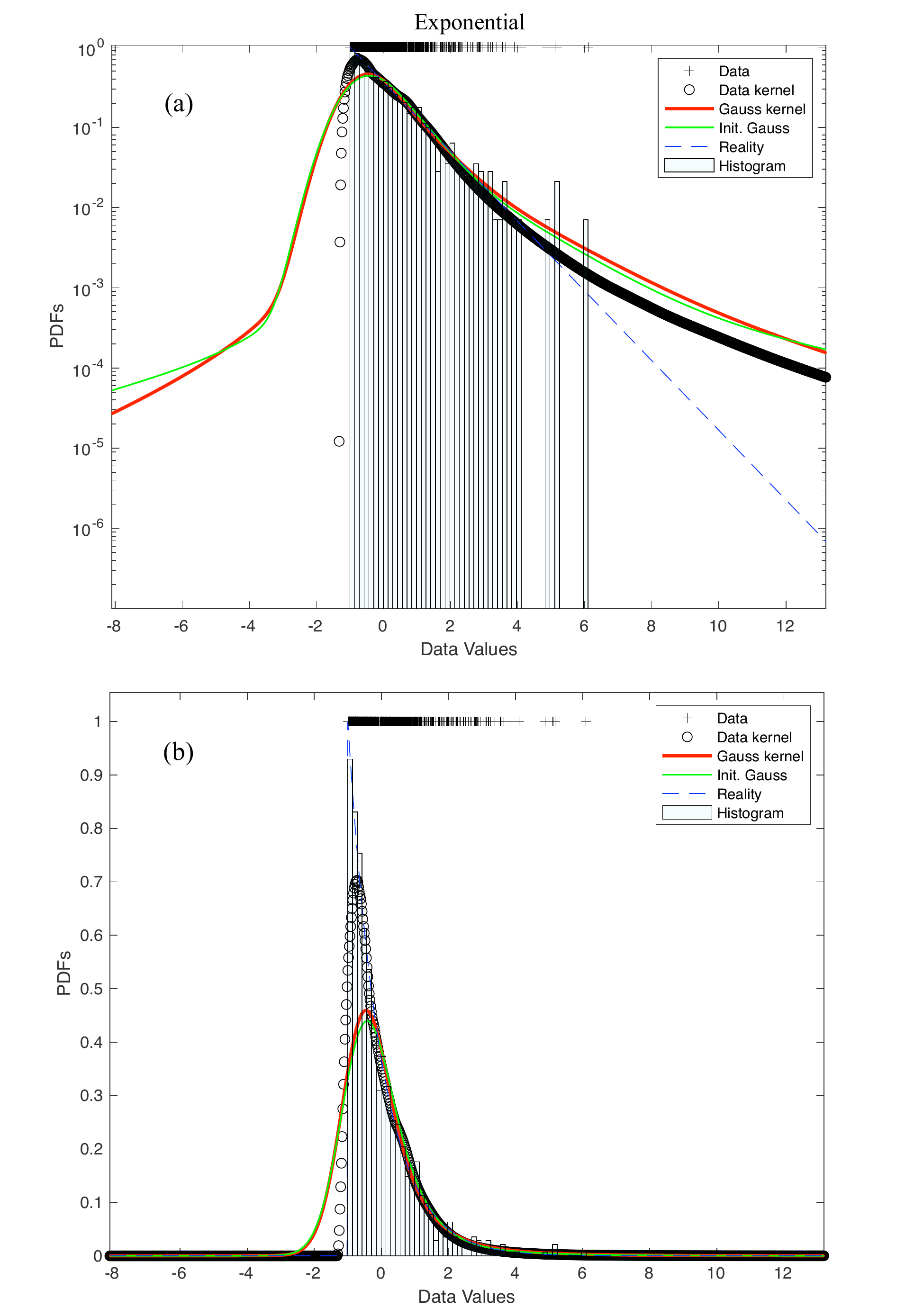}
 \caption{a) Semilog and b) linear plots of iteratively estimated densities for a single realization of 1,000 shifted Exponential data points using data-based kernel (black symbols) and Gaussian-based kernels (red curves).  Also shown are the single-pass Gaussian kernel estimate (green curves) and 20-bin histograms.  Blue dashed line is underlying ``true'' density function.}
 \label{fig:Exp_1000}
 \end{figure}

{Next,} we use a shifted exponential (the arbitrary shift is added to ensure functionality of the code) with density function 
\bl
\begin{equation*}
f(x)= \begin{cases}
\frac{1}{\sigma} \exp \left (-\frac{x-\mu+1/\sigma}{\sigma}\right ), & \text{for} \ x\ge\mu-1/\sigma\\
 0, & \text{else}.
 \end{cases}
 \end{equation*}
 \el
This density has arbitrary mean $\mu$ and variance $1/\sigma^2$.  In the plots that follow we set $\mu=0$ and $\sigma=1$.  For this skewed density it becomes clear that a symmetric (Gaussian in this case) kernel is not an effective interpolant (Fig. \ref{fig:Exp_1000}).  While \cite{Silverman1986} suggests using a skewed (say, lognormal) kernel for this kind of data, our method does not rely on interpretation and user intervention.  And while a Gaussian kernel is not particularly useful for this kind of data, the iterated data-based kernel typically has MISE of about half that of the Gaussian (Table \ref{tab:rmse}).  Because the underlying optimal $h_0$ is much smaller than that estimated from assuming a Gaussian kernel, the iterations do not converge (and in fact tend to diverge) until $h_0$ decreases several times, requiring on the order of 30 or more iterations for 1,000 data points. 

\subsection{Cauchy data}
\begin{figure}
 \centering
 \includegraphics[width=5.5in]{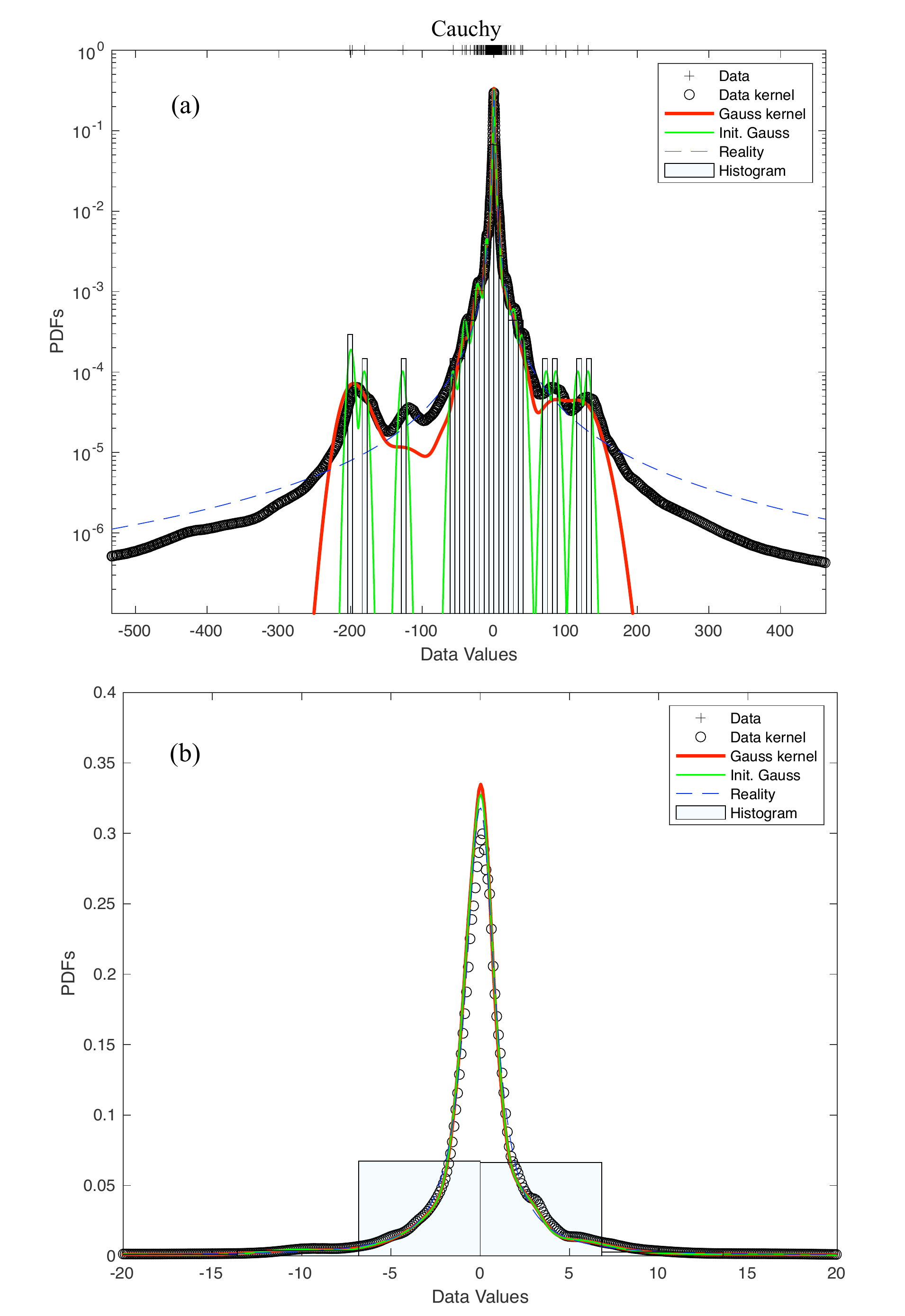}
 \caption{a) Semilog and b) linear plots of iteratively estimated densities for a single realization of 1,000 Cauchy data points using data-based kernel (black symbols) and Gaussian-based kernels (red curves).  Also shown are the single-pass Gaussian kernel estimate (green curves) and uniform 20-bin histograms.  Blue dashed line is underlying ``true'' density function.}
 \label{fig:Cauchy_1000}
 \end{figure}

Heavy-tailed data present a problem for kernel density estimation because of the extremes that may accompany the data.  This leads to very wide spacing between extreme data points and difficulty interpolating the density here.  This also means that the $x$-discretization of the kernel must use a large number of points in order to represent the near-origin ``spikiness'' of the density as well as the very long range.  The existence of one or two super-extreme values can lead to numerical problems.  In our 100-realization ensemble of 1,000 Cauchy data points, two realizations failed to converge in 100 iterations with the data-based kernel because of data values in the 50,000 range.  A typical realization {shows} that the converged data-based kernel tends to both interpolate between, and extrapolate beyond, extreme values better than the Gaussian kernel, but still represents the fine-scale near the origin where most of the data reside (Fig. \ref{fig:Cauchy_1000}).  In the ensemble, the MISE estimated for the data-based kernel is {less} than for the Gaussian kernels (Table \ref{tab:rmse}), but there is substantial variability that we did not track in these numbers (and which is {least}) between realizations.

\subsection{Maximally-skewed $\alpha$-stable data}
\begin{figure}
 \centering
 \includegraphics[width=5.5in]{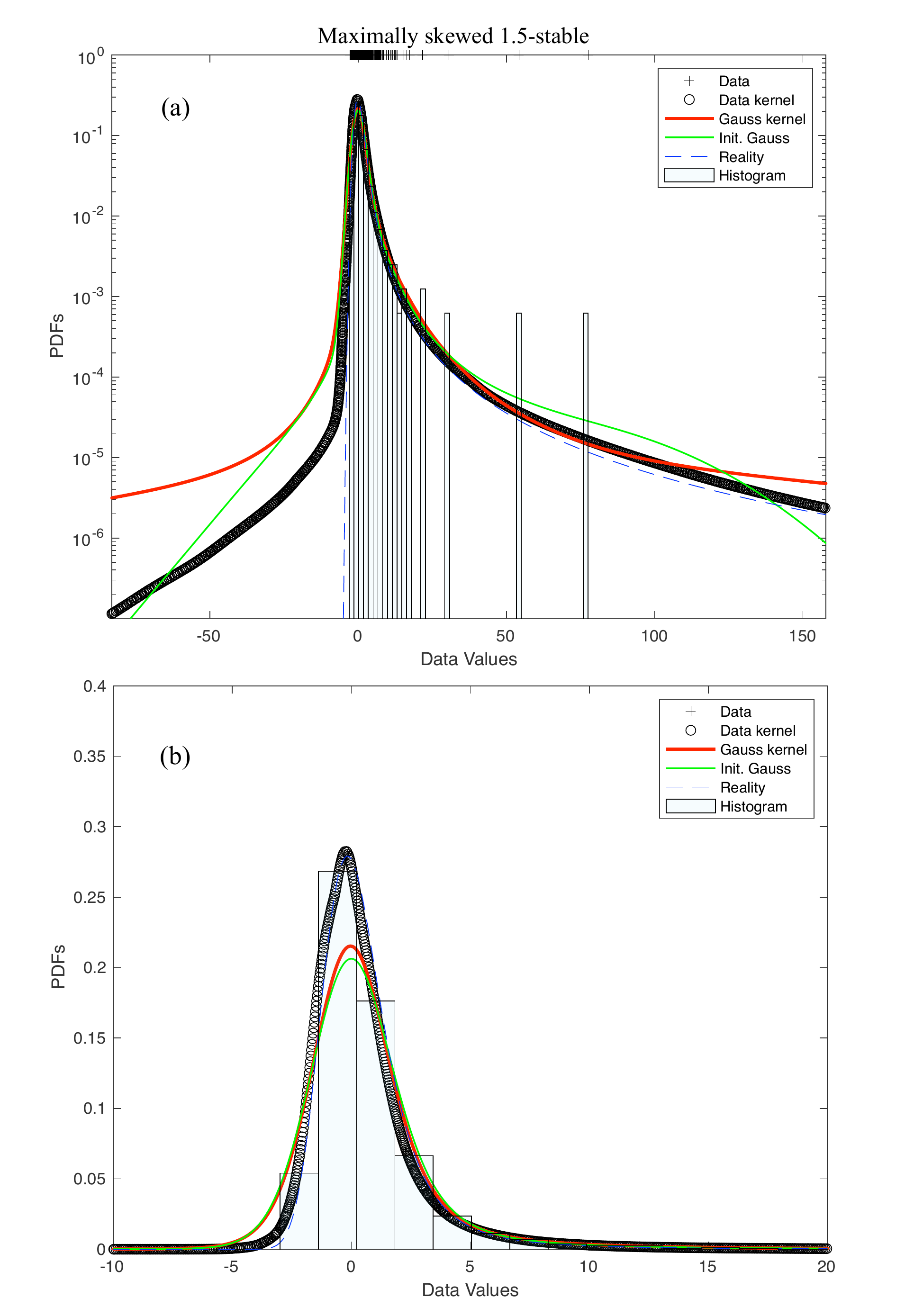}
 \caption{a) Semilog and b) linear plots of iteratively estimated densities for a single realization of 1,000 maximally-skewed, 1.5-stable data points using data-based kernel (black symbols) and Gaussian-based kernels (red curves).  Also shown are the single-pass Gaussian kernel estimate (green curves) and uniform 20-bin histograms.  Blue dashed line is underlying ``true'' density function.  \label{fig:Stable_1000} }

 \end{figure}

Stable random variables are characterized (among many other ways) as those to which sums of IID random variables converge \citep{Samorod_book}.  Finite-variance random variables converge to (are in the domain of attraction of) a Gaussian, which is itself a stable RV.  When, for some constant $0<\alpha<2$, only those moments of order $\alpha$ and greater are infinite (such as for Pareto (power-law) distributed RVs), then those RVs are in the domain of attraction of a $\alpha$-stable.  These RVs {arise} in hydrology quite naturally, because they describe waiting times that a particle might take when trapped in a sequence of fractal immobile zones (see \cite{Schumer2003,Benson_review}).  Depending on the skewness parameter, one or both of the tails of a $\alpha$-stable decay like $\sim |x|^{-1-\alpha}$; therefore all moments of order $\alpha$ and greater diverge.  The density is only expressible in closed-form for a few instances, but most statistical packages will readily calculate the density to any desired tolerance and generate sequences of the random variable. Here, we choose a maximally-skewed, standard 1.5-stable for analysis, using the parameterization in the MATLAB statistics package (also called the 0-parameterization in \cite{Nolan_online}).  
The ensemble MISE for the data-based kernel is about 1/5 that of the iterated Gaussian kernel, suggesting that both the heavy-tailed and skewed nature of this example is especially well-suited to the proposed method (Fig. \ref{fig:Stable_1000}).

\subsection{Breakthrough data}
The creation of ``breakthrough curves'' from particle-tracking simulations is a tricky proposition.  Classically, histograms are used, which means manually choosing either constant or variable bin sizes and locations.  The variance of the estimated density is inversely proportional to bin size, total number of particles, and the estimated concentration \citep{Chakraborty}, and the histogram-based density is discontinuous and may frequently be zero when particle arrival times are widely separated, especially in the late-time tail. The zeros make comparison to non-zero data difficult, so several methods are typically used to create a non-zero PDF.

For example, one may construct an empirical cumulative distribution (ECDF) function that is strictly increasing by several means and then make a non-zero empirical PDF using finite differences on the ECDF.  For example, order particle arrival times of $N$ particles $T_1, T_2,...T_N$ and at each point the $\ECDF(T_i) = i/N$.  Then the empirical PDF is $\mathrm{EPDF}((T_{i+1}+T{i})/2) = (\ECDF(T_{i+1})-\ECDF(T_i))/(T_{i+1}-T_i); i=1..N-1$.  The ECDF can also use a regularly spaced time grid and count numbers of particles arriving between grid points, and grid points without arrivals are neglected, once again giving a strictly increasing ECDF.  In this example, we compare these methods to the iterative kernel-based techniques developed here.  

\begin{figure}
 \centering
 \includegraphics[width=5.5in]{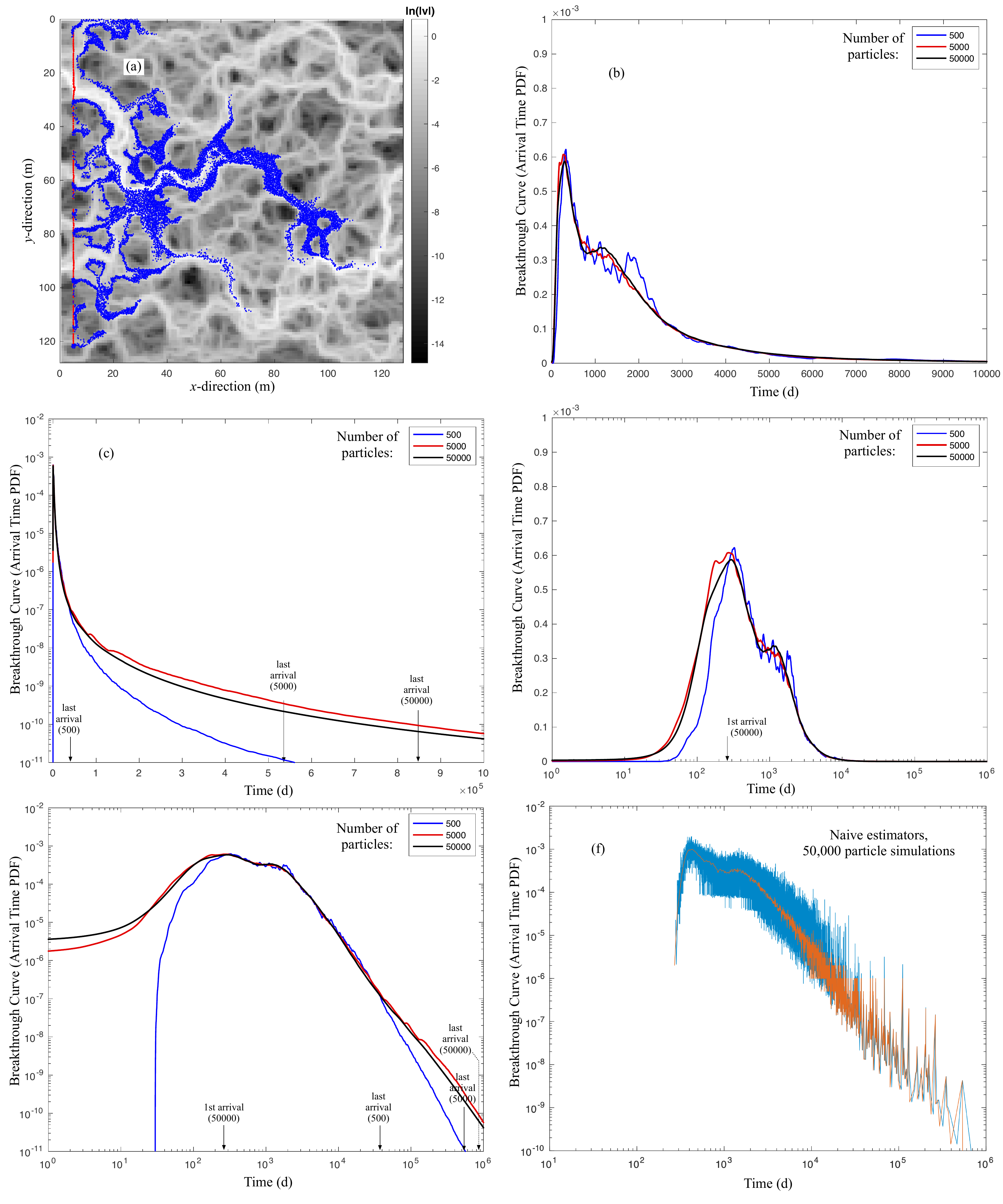}
 \caption{a) 50,000-particle simulation positions at $t=1$ day (red) and 250 days (blue) on a gray-scale quilt of log-velocity magnitude.  Mean flow is left-to-right.  b) through d) Plots of estimated arrival-time densities (breakthrough curves) using differently scaled axes for 500, 5,000, and 50,000 particle simulations.   The densities are estimated using the data-based kernel.   }
 \label{fig:v_field}
 \end{figure}

For particle arrival times, we solved the steady-state groundwater flow equation $\nabla \cdot K \nabla h = 0$ in 2-D using finite-differences on a square 128 $\times$ 128 m grid with constant grid discretization of 1 $\times$ 1 m (Figure \ref{fig:v_field}a).  The hydraulic conductivity $K$ is a log-Normal random variable with $E(\ln(K))=1$ and $VAR(\ln(K))=16$ and an exponential autocorrelation function for $\ln(K)$ with correlation length of 5 m.  The left and right boundaries $x=0$ and $x=128$ are Dirichlet with $h=1$ and $h=0$, respectively.  The top and bottom boundaries $y=0$ and $y=128$ are Neumann with $\partial h /\partial y =0$.  The resulting velocities take the solved $h$ field and apply $v = -K\nabla h$, once again using finite-differences. These velocities vary in magnitude from about $3 \times 10^{-7}$ to 2.1 m/d (Figure \ref{fig:v_field}a).  Particles are placed in a line near the left boundary and each particle's position vector tracked via a discretized Ito equation $X(t+\Delta t) = X(t) + (v + \nabla\cdot (D)) + \sqrt{2\Delta t}B \mathcal{N}$, where $D = (D_m + A_T |v|) I +(A_L-A_T) vv^T/|v|$  is a dispersion tensor that has a decomposition $D=BB^T$, $\mathcal{N}$ is an independent 2-D standard normal vector, and $D_m = 8 \times 10^{-5}$ m$^2$/d is molecular diffusion, $A_T = 10^{-3}$ m is transverse dispersivity, and $A_L = 5 \times 10^{-3}$ m is longitudinal dispersivity.  The number of particles placed in any cell is proportional to the velocity magnitude in that cell (i.e., a flux-weighted source). A plot of particle positions at elapsed times of 1 and 250 days (just before arrival of first particle at the right-hand side) suggests the wide range of arrival times that can be expected.  We ran simulations using 500, 5000, and 50000 particles to judge the efficacy of density estimates.

The data-based kernel estimates are remarkably similar on a linear plot (Figure \ref{fig:v_field}b).  Two estimates of the EPDF using 50,000-particle arrival times (Figure \ref{fig:v_field}f) show considerable noise at late time due to wide separation of late particles.  They also do not have density weight before the first particle arrival because the ECDF is zero for the early times.  This is a commonly accepted feature, but as the number of particles become larger (or goes to infinity in the case of kernel density estimates) the empirical density of early arrivals should grow.  The kernel density estimates allow for this (having a calculation domain as large as desired).  Because the estimates of global bandwidth $h_0$ are larger for smaller $n$, the 500-particle density estimates have a greater probability at early time, which can be identified using logarithmic time axes (Figure \ref{fig:v_field}d-e).  The late-time tail estimates are close for all three particle numbers ((Figure \ref{fig:v_field}e) until some time after the final particle arrival time in the 500-particle simulation, when that tail starts to drop off a fair amount.  Overall, it is fair to say that the 5,000-particle simulation gives similar enough results to the 50,000-particle simulation that the latter is superfluous.
 
A serious problem with the kernel density estimates is that the time grid along which the density (hence kernel for subsequent iterations) needs to be quite large.  The density is ``spiky'' enough to warrant a grid size of about 20 days or less, and the last particle arrive on the order of $10^6$ days, so a linearly partitioned grid is a vector on the order of 50,000 to 100,000 elements, making the convolutions quite slow.  Convergence is also slow because of the high skewness, so the estimates are computationally expensive.  Because of this we looked at two alternatives: 1) use of a log-spaced discretization grid (which was used to generate Figure \ref{fig:v_field}, and 2) the {\em ad-hoc} correction of \cite{Pedretti_kernel}, which is detailed immediately.

\subsection{Universal Adaptive Bandwidth of \citet{Pedretti_kernel}}
\begin{figure}
 \centering
 \includegraphics[width=6in]{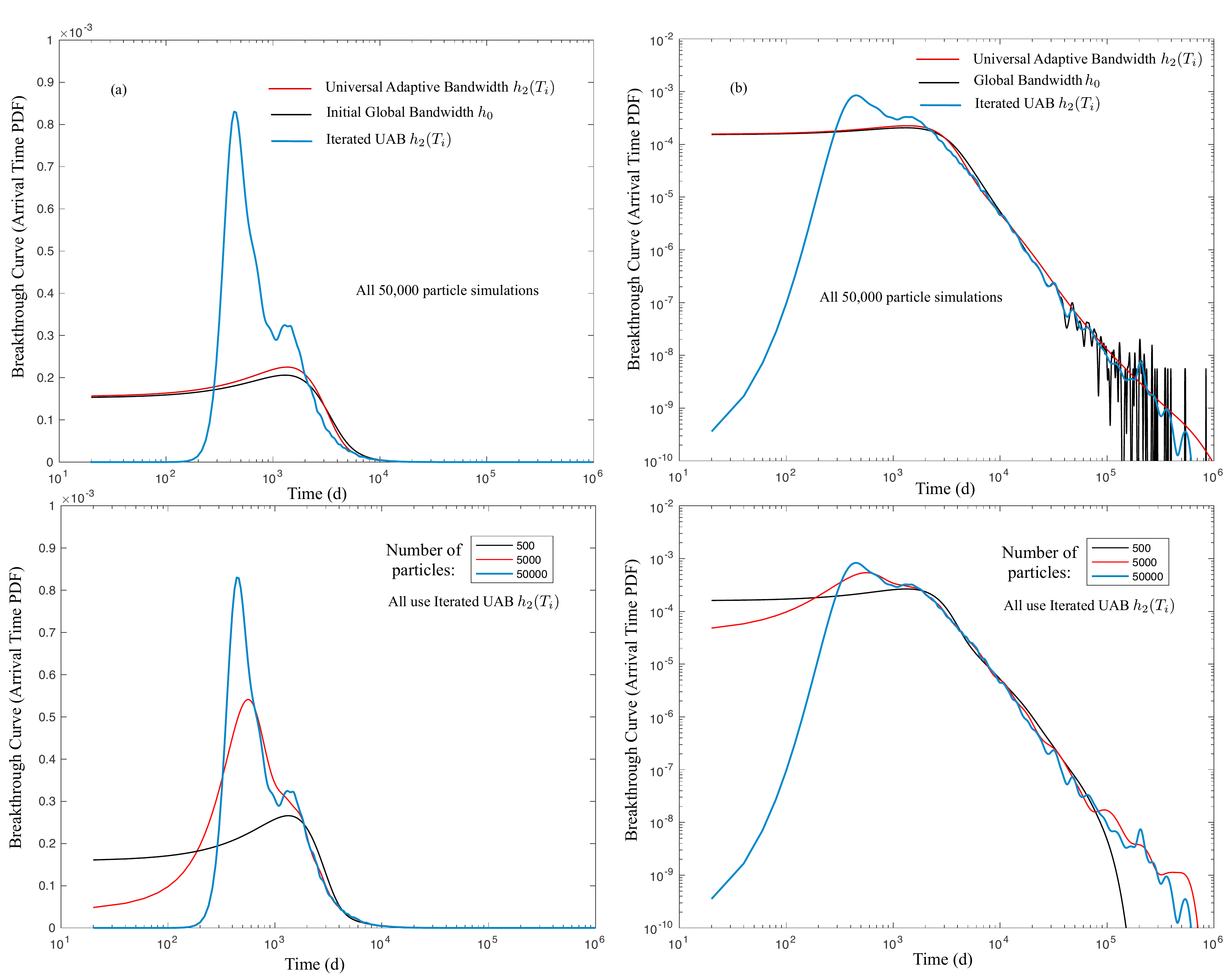}
 \caption{a) Semi-log and b) Log-log plots of estimated arrival-time densities (breakthrough curves) and the 50,000 particle simulation also shown in Fig. \ref{fig:v_field}.  Red curves use the universal adaptive bandwidth of \cite{Pedretti_kernel} (our eq. \eqref{eq:h2}, black curves use the single-valued global bandwidth $h_0$ from eq. \eqref{eq:eps_Gauss}.  Blue curves show our suggested procedure of iteratively re-estimating $h_0$ (using the current iteration of the density function) using eq. \eqref{eq:h_0} and plugging into eq.\eqref{eq:h2}.  c) and d) are semi-log and log-log plots of the iterative UAB method for 50,000, 5,000, and 500-particle simulations, illustrating the need for large particle numbers.}
 \label{fig:Pedretti}
 \end{figure}

These authors recognize that early arrival tails are often much thinner than late-arrival tails and seek to adjust the bandwidth assigned to early versus late data accordingly.  They suggest constructing the $\ECDF(T_i)$ at each data point and constructing a variable bandwidth at each point by taking a weighted average of the single global bandwidth $h_0$ and the classical adaptive bandwidth:
\bl
\begin{equation}\label{eq:h2}
h_2(T_i) = (1-\ECDF(T_i))h_0 + \ECDF(T_i)\times h_i,
\end{equation} 
\el
where $h_2$ is their ``universal global bandwidth'' (UAB), and $h_i$ is the adaptive bandwidth given in eq. \eqref{eq:adapt}.  \citet{Pedretti_kernel} choose a standard Gaussian kernel in their paper so we do the same here.  We use our algorithms to estimate $h_0$ and find that the numerical estimate gives $h_0 =$ 2,370 days and the assumption of Gaussian data gives  $h_0 =$ 1,432 days.  Both give similar results and we choose the smaller value to give slightly better visual PDFs.  We also used the same 50,000 particle arrival times as in the previous section.  Application of eq. \eqref{eq:h2} results in visually acceptable smoothing of the late-time tail (Fig. \ref{fig:Pedretti}a) relative to the constant bandwidth $h_0$.  An adaptive kernel using $h_i$ gives essentially identical results to the UAB in the late time tail and similar results for small time (Fig. \ref{fig:Pedretti}a). The small time estimate of the density is clearly over-smoothed (Fig. \ref{fig:Pedretti}b) because the values of the bandwidth there approach $h_0=$ 1,432 d, which is a large bandwidth considering the first arrival time of 277 days in the 50,000 particle simulation.  This large bandwidth gives substantial density in negative time.  It is important to note that we used our estimates of $h_0$, while \cite{Pedretti_kernel} use a method that estimates the integrals in eq. \eqref{eq:h_0}.  This may explain the difference in their effectiveness of the UAB, and also suggests a potentially improved method: The numerically estimated $h_0$ using data appears to give too large a value; therefore, a smoothed version of the density is likely to have a smaller integrated second derivative in eq. \eqref{eq:h_0}. So after each iteration, the estimated density can be used to estimate $\int f^{\prime \prime}(x)dx$ by finite differences.  This $h_0$ is used in the UAB (eq. \eqref{eq:h2}) until closure is reached.  Closure is expected because smaller $h_0$ will give larger $\int f^{\prime \prime}(x) dx$, so the two factors will eventually counterbalance. Indeed, closure occurs for the 50,000-particle interpolation problem in about 15 iterations, and yields a density that conforms much more closely to our previous results based on a data-based kernel (compare blue curves in Figs. \ref{fig:Pedretti} a,b to \ref{fig:v_field} d,e). However, this procedure requires that the estimated density is a copy of the real density, which is most applicable for large data numbers (and will be true only as $n\rightarrow \infty$), so we check the 5,000 and 500 particle simulations relative the 50,000 (Figs. \ref{fig:Pedretti} c and d)).  Clearly, estimating the curvature properties of the ``real'' density using a smoothed version does not perform well for smaller data sets.

 \section{Conclusions and Recommendations}\label{sec:conclusion}
The application of an optimal, iterative algorithm for kernel density estimation (using an evolving, data-based, kernel) is possible with a few caveats.  First, because the underlying data density is, in general, unknown, a method is needed to estimate a MISE-minimizing global bandwidth $h_0$.  We show that a Fourier-transform based method can obtain an unbiased estimate for any kernel, and the exact value for a Gaussian kernel.  This Gaussian kernel ``starts'' the new algorithm by generating a first continuous estimate of the density.  This density is then used to construct the kernel for subsequent density estimates.  Second, creating a ``standard'' kernel based on the current density estimate requires an estimate of the scale parameter of the density.  We use a value based on the interquartile range divided by 1.5.  This value is intermediate for several known densities and works well for a range of known densities. Third, because the final iterated version does not use a Gaussian kernel, the initial estimate of $h_0$ will be in error.  We show that for some common densities, the naive estimate of $h_0$ will err on the large side.  Furthermore, we show for a wide range of densities that the estimate of $h_0 \sim n^{-\gamma}$, with $\gamma$ being a minimum for Gaussian data and increasing systematically as the tails become heavier (including exponential and power-law).  Therefore, the iterative scheme allows $h_0$ to decrease if the algorithm fails to demonstrate convergence.  As expected, for Gaussian data, the data-based kernel converges rapidly to a form similar to that given by the Gaussian kernel.  For skewed and/or heavy-tailed data, convergence is slower and only occurs when $h_0$ is allowed to decrease toward its actual, optimal value.  Overall, the data-based kernel gives significantly smaller ensemble MISE than either 1) an iterated (adaptive bandwidth) Gaussian kernel, 2) a single-pass adaptive bandwidth Gaussian kernel, and 3) a single-pass Gaussian kernel with a single global value of $h_0$.  Furthermore, the iterative algorithm can be extended to traditional estimation techniques using known (say, Gaussian) kernels by re-estimating $h_0$ using intermediate density estimates. This assumes that the curvature of the interpolated density is a good estimate of the curvature of the ``real'' density. Using the the method of \cite{Pedretti_kernel}, we show that a large number of data are needed.

The derivation of the optimal kernel and global bandwidth solved a minimization problem for one variable $h_0$ based on kernel shape and the Fourier transform of actual data (Appendix C).  A more difficult problem of optimizing a separate $h_i$ for each data point may be possible.  This would eliminate the potentially dubious Taylor-series-based assumptions of the power-law weighting scheme used in Eq. \eqref{eq:adapt} to adjust each data point's bandwidth.  We leave this for a future paper.

\section{Acknowledgements}
This material is based upon work supported by, or in part by, the US Army Research Office under Contract/Grant number W911NF-18-1-0338. The authors were also supported by the National Science Foundation under awards EAR-1417145, DMS-1614586, DMS-1911145,  EAR-1351625, EAR-1417264, EAR-1446236, and CBET-1705770.   Matlab codes for generating all results in this paper are held in the public repository \url{https://github.com/dbenson5225/kernel-density-estimation}.

\section{Appendix A: Mathematical Background}
The idea of optimal global bandwidth \citep{Silverman1986} stems from using a truncated Taylor series to represent the terms in the MISE.  We begin with the fact that the expectation of the density estimate constructed from a set of independent observations is the sum of the expectations of the weights associated with each observation so that
\bl
\beq
\mathbb{E} [\bar f(x)] = \frac{1}{n} \sum_{j=1}^n \mathbb{E}\left [ \frac1hK\biggl( \frac{x-X_j}{h} \biggr) \right ] = \int \frac1h K \biggl( \frac {x-\xi} {h} \biggr) f(\xi) d\xi. 
\eeq
\el
\noindent Similarly, we compute the variance as
\bl
\begin{align*}
\text {Var} [\bar {f}(x) ]&= \text{Var} \left [\sum_{j=1}^n \frac{1}{nh} K\biggl(\frac {x-X_j}{h} \biggr) \right ]= \sum_{j=1}^n \frac{1}{n^2} \text{Var} \left [\frac1h K\biggl(\frac {x-X_j}{h} \biggr) \right ] \\
&=\frac1n \int \left( \frac{1}{h} K \biggl( \frac {x-\xi} {h} \biggr) \right)^2 f(\xi) d\xi - \frac1n \left( \int \frac{1}{h} K \biggl( \frac {x-\xi} {h} \biggr) f(\xi) d\xi \right)^2. 
\end{align*}
\el
\noindent The bias at any point is
\bl
\begin{align*}
B(x)&=\mathbb{E} [\bar f(x) ]-f(x)=\int\frac1h K \biggl( \frac{x-\xi}{h} \biggr) f(\xi) d\xi -f(x) \\
&=\int K(z)f(x-hz)dz  -f(x) \\
&=\int K(z)(f(x-hz)-f(x))dz.
\end{align*}
\el
\noindent With this, the MISE is written as
\bl
\beq
\label{MISEBV}
\left .
\begin{split}
\text{MISE}&=\int B(x) ^2 dx +\int \text{Var} [ \bar f(x)] dx \\
&= \int \biggl(  \int K(z)(f(x-hz)-f(x))  \biggr)^2 dz dx \\
& +\frac1n  \left(   \iint \frac{1}{h^2} K \biggl( \frac{x-\xi}{h}  \biggr)^2 f(\xi) d\xi  dx - \int  \left( \int \frac{1}{h} K \biggl( \frac {x-\xi} {h} \biggr) f(\xi) d\xi \right)^2 dx \right).
\end{split}
\right \}
\eeq
\el
\noindent The bias contribution is simply the effect of the kernel smoothing on the real density, which does not depend on the sample size $n$.  The variance obviously grows smaller as $n$ increases and (not completely obviously) as $h$ increases.  This expression is difficult to minimize exactly, although for both $K$ and $f$ Gaussian, the convolutions yield Gaussians and an exact result may be computed \citep{Silverman1986}. The vast majority of work done with KDE is to use asymptotic expansions of certain functions, with some questionable assumptions regarding their validity and application.  
For example, the density at $x-hz$ is typically approximated for $hz \rightarrow 0$, even though the goal is to find a finite $h$ and $z$ may be arbitrarily large in the integral.  Still, using a truncated Taylor series, namely
\bl
\begin{equation*}f(x-hz)=f(x)-hz f^\prime(x) + \frac12 h^2z^2 f^{\prime\prime}(x) + \mathcal{O}(h^3)
\end{equation*}
\el
gives
\bl
\begin{align*}
B(x)&=-hf^\prime (x) \int z K(z) dz + \frac12 h^2 f^{\prime \prime}(x) \int z^2 K(z) dz + \mathcal{O}(h^3) \\
&= -hf^\prime (x) \mu_1(K) + \frac12 h^2 f^{\prime \prime}(x) \mu_2(K) + \mathcal{O}(h^3),
\end{align*}
\el
\noindent where $\mu_n(K)$ denotes the $n^{th}$ moment of the kernel.  Clearly, using a zero-mean (i.e., symmetric or properly shifted) kernel eliminates the first term on the RHS, and indeed, letting $h\rightarrow 0$ eliminates bias altogether, but at the cost of increasing the noise in the estimate. Assuming a finite mean and proper shifting, the squared bias is simply (after truncation of higher-order terms)
\bl
\begin{equation*}
\int B(x)^2 dx \approx \frac14 (h^2 \mu_2(K))^2 \int (f^{\prime\prime})^2 dx.
\end{equation*}
\el
\cite{Silverman1986} uses the bias approximation, the substitution $z=(x-\xi)/h$, and another application of Taylor series to reduce the local variance term to 
\bl
\begin{align*}
\text{Var}[\bar f(x)]&\approx \frac{1}{nh} \int K(z)^2 f(x-zh) dz - \frac1n \left (f(x)+\mathcal{O}(h^2) \right)^2 \\ 
&\approx \frac{1}{nh} \int K(z)^2 \left (f(x)-hzf^\prime(x) +\mathcal{O}(h^2) \right) dz +\mathcal{O}(n^{-1}) \\ 
&\approx\frac{1}{nh} f(x) \int K(z)^2  dz,
\end{align*}
\el
\noindent which, when integrated over $x$ yields
\bl
\begin{equation*}
\int\text{Var}[\bar f(x)]  {dx} \approx \frac{1}{nh} \int K(z)^2  dz.
\end{equation*}
\el
All told, this gives a MISE of
\bl
\begin{equation*}
\text{MISE} \approx \frac14 (h^2 \mu_2(K))^2 \int f^{\prime\prime}(x)^2 dx +\frac{1}{nh} \int K(z)^2  dz,
\end{equation*}
\el
based on the assumptions of small $hx$, large $n$, and $n \gg1/h^2$, all of which are likely to be bad assumptions in practice.  Taking $d(\text{MISE})/dh$ and setting this expression to zero clearly gives the global estimate eq. \eqref{eq:h_0} in one-dimension. 

\section{Appendix B: Global Bandwidth Estimation Using Fourier Methods}
In this section, we introduce a new method using the Fourier transform that will allow us to create an unbiased estimator for the MISE and minimize this function in order to select the optimal bandwidth $h$.
Throughout, we will use the form of the transform common to fast Fourier transform routines, namely
\bl
\begin{equation*}
\hat g(\omega) = \int e^{-2\pi i \omega x} g(x)dx
\end{equation*}
\el
for any sufficiently smooth function $g(x)$.
Recall that the MISE can be written as the sum of a bias and variance term as in \eqref{MISEBV} so that
\bl
\beq
\label{MISE1}
\MISE = \int \bias(x)^2 \ dx + \int \var [ \bar{f}(x)] \ dx
\eeq
\el
where the bias is
\bl
\begin{equation*}
\bias(x) = \int K(z) f(x-hz) \ dz - f(x)
\end{equation*}
\el
and the variance is given by
\bl
\begin{equation*}
\var [\bar{f}(x)] = \frac{1}{n} \left [ \int \frac{1}{h^2} K \left (\frac{x-z}{h} \right )^2 f(z) \ dz - \left ( \int \frac{1}{h} K \left (\frac{x-z}{h}\right ) f(z) \ dz\right )^2 \right ].
\end{equation*}
\el
Using Fourier methods, we first compute the bias term. 
Taking the transform of the bias and making the change of variables $y = x-hz$, we find
\begin{eqnarray*}
\mcB(\omega ) & = & \iint K(z) f(x- hz) e^{-2\pi i x\omega} \ dz dx - \mcF(\omega )\\
& = & \iint K(z) f(y) e^{-2\pi i \omega(y + hz)} \ dy dz - \mcF(\omega )\\
& = & \left (\int K(z) e^{-2\pi i \omega hz} \ dz - 1 \right ) \mcF(\omega )\\
& = & \left ( \mcK(h\omega) - 1 \right ) \mcF(\omega).
\end{eqnarray*}
Therefore, by Plancherel's Theorem, the bias term in equation \eqref{MISE1} becomes
\bl
\begin{equation*}  \int \bias(x)^2 \ dx =  \int \mcB(\omega )^2 \ d\omega  = \int \left ( \mcK(h\omega ) - 1 \right )^2 \mcF(\omega )^2 \ d\omega .
\end{equation*}
\el

To compute the associated variance term in equation \eqref{MISE1}, we first split it into two parts so that
\bl
\begin{equation*}
\int \var [ \bar{f}(x)]  dx= \frac{1}{n} \left ( I - II \right ).
\end{equation*}
\el
The first term is then
\bl
\begin{equation*}
I = \frac{1}{h^2} \iint K \left ( \frac{x-z}{h} \right )^2 f(z) \ dz dx
\end{equation*}
\el
and satisfies
\bl
\begin{eqnarray*}
I  & = & \frac{1}{h} \iint K(y)^2 f(x - hy) \ dy dx\\
& = & \frac{1}{h} \int K(y)^2 \left ( \int f(x - hy) dx \right ) dy\\
& = & \frac{1}{h} \left (\int K(y)^2 \ dy \right ) \left ( \int f(\xi) d\xi \right )\\
& = & \frac{1}{h}\int K(y)^2 \ dy
\end{eqnarray*}
\el
due to the change of variables $y = (x-z)/h$ and then $\xi = x-hy$, as well as the fact that $f(x)$ is a pdf.
To compute $II$, we write it as
\bl
\begin{equation*}
II = \int P(x)^2 \ dx
\end{equation*}
\el
where
\bl
\begin{equation*}
P(x) =  \frac{1}{h}\int  K \left (\frac{x-z}{h}\right ) f(z) \ dz =  \int K(y) f(x-hy) \ dy.
\end{equation*}
\el
Of course, the transform of $P(x)$ has already been identified in the computation of the integrated bias term.
In particular, it is given by
\bl
\begin{equation*}
\mcP(\omega ) = \mcK(h\omega) \mcF(\omega).
\end{equation*}
\el
Using Plancherel's theorem as before, we find
\bl
\begin{equation*}
II = \int \mcP(\omega)^2 \ d\omega = \int \mcK(h\omega)^2 \mcF(\omega)^2 \ d\omega.
\end{equation*}
\el
With this Fourier representation of the bias and variance integrals, we may explicitly write the MISE in terms of integrals of transformed functions, namely
\bl
\beq
\label{eq:FTMISE}
\MISE_n(h) = \frac{1}{nh} \int \mcK(\omega)^2 \ d\omega + \int \left ( (\mcK(h\omega)-1)^2- \frac{1}{n} \mcK(h\omega)^2 \right ) \mcF(\omega)^2 \ d\omega.
\eeq
\el
Note that we have used $\int K(y)^2 \ dy = \int \hat{K}(\omega)^2 \ d\omega$ in the first term to write the MISE depending upon $\hat{K}$ rather than $K$.

Unfortunately, this expression still requires knowledge of the Fourier transform, $\mcF(\omega)$, of the unknown pdf and thus cannot be used to choose the optimal bandwidth $h$.
Instead, we will rely on an empirical distribution to approximate $f$, and thus $\mcF$.
Given $n$ observations of the distribution $f(x)$, which are denoted $X_1, ..., X_n$, we define the empirical (or observed) distribution
\bl
\begin{equation*}
f_n(x) = \frac{1}{n} \sum_{j=1}^n \delta(x - X_j)
\end{equation*}
\el
so that the corresponding transform of this function is
\bl
\beq\label{eq:FTf}
\mcF_n(\omega) = \frac{1}{n} \sum_{j=1}^n \int e^{-2\pi i x \omega} \delta(x - X_j) \ dx =  \frac{1}{n} \sum_{j=1}^n e^{-2\pi i \omega X_j} .
\eeq
\el
Now, as $n \to \infty$, we find $f_n \to f$ and $\mcF_n \to \mcF$. In fact, we have an asymptotic estimate for the expected value of $\mcF_n$, which implies
\bl
\be
\label{EFn}
\E \left [\mcF_n(\omega)^2 \right ] \approx \left (1 - \frac{1}{n} \right ) \mcF(\omega)^2 + \frac{1}{n}
\ee
\el
as $n \to \infty$.

Therefore, by using the empirical distribution, we can define and utilize an unbiased estimator for the $\MISE$.
For fixed $n \in \mathbb{N}$ and any $h \geq 0$, define
\bl
\beq
\begin{split}
\label{eq:epsdef}
\mcE_n(h) &= \frac{2}{n}\int K(h\omega) d\omega + \int \left [ \left ( 1 -\frac{1}{n} \right ) \mcK(h\omega)^2 - 2\mcK(h\omega) \right ] \mcF_n(\omega)^2 \ d\omega \\
&= \frac{2}{nh} K(0) + \int \left [ \left ( 1 -\frac{1}{n} \right ) \mcK(h\omega)^2 - 2\mcK(h\omega) \right ] \mcF_n(\omega)^2 \ d\omega.
\end{split}
\eeq
\el
Then, $\mcE_n(h)$ and $\MISE_n(h)$ must attain their minimum values at the same $h$.
Therefore, given a sample $X_1, ..., X_n$ of $n$ draws from $f(x)$, we define the optimal bandwidth by
\bl
\begin{equation*}
h_\mcE = \argmin\limits_{h \geq 0} \ \mcE_n(h).
\end{equation*}
\el
Computationally approximating the global bandwidth using this value of $h_\mcE$ is instrumental to the algorithm proposed in Section $2$.

Finally, we justify the claim that $\mcE_n(h)$ is an unbiased estimator of the $\MISE$.
We first note that by the Fourier inversion property we have
\bl
\begin{equation*}
\frac{1}{h}K(0) = \frac{1}{h} \int \mcK(\omega) e^0 \ d\omega  = \int \mcK(h\omega) \ d\omega.
\end{equation*}
\el
Then, taking the expectation of $\mcE_n(h)$ and inserting the convergence result \eqref{EFn}, we find
\bl
\begin{eqnarray*}
\E \left [\mcE_n(h) \right ] & = & \frac{2}{n}\int \mcK(h\omega) d\omega + \int \left [ \left ( 1 -\frac{1}{n} \right ) \mcK(h\omega)^2 - 2\mcK(h\omega) \right ] \E \left[ \mcF_n(\omega)^2 \right ] \ d\omega\\
& = & \frac{2}{n}\int \mcK(h\omega) d\omega + \int \left [ \left ( 1 -\frac{1}{n} \right ) \mcK(h\omega)^2 - 2\mcK(h\omega) \right ] \left ( \left (1 - \frac{1}{n} \right ) \mcF(\omega)^2 + \frac{1}{n} \right ) \ d\omega\\
& = & \left (1 -\frac{1}{n} \right ) \int \mcK(h\omega)^2 \left ( \left (1 - \frac{1}{n} \right ) \mcF(\omega)^2 + \frac{1}{n} \right ) \ d\omega - 2\left (1 -\frac{1}{n} \right ) \int \mcK(h\omega) \mcF(\omega)^2 \ d\omega \\
& = & \left (1 -\frac{1}{n} \right ) \int \left [\mcK(h\omega)^2 \left ( \left (1 - \frac{1}{n} \right ) \mcF(\omega)^2 + \frac{1}{n} \right ) - 2 \mcK(h\omega) \mcF(\omega)^2 \right ]\ d\omega \\
& = & \left (1 -\frac{1}{n} \right ) \int \left [ \left (\mcK(h\omega)^2 - 2 \mcK(h\omega)  - \frac{1}{n} \mcK(h\omega)^2 \right ) \mcF(\omega)^2 + \frac{1}{n} \mcK(h\omega)^2 \right ]\ d\omega \\
& = & \left (1 -\frac{1}{n} \right ) \left [\int \left ( (\mcK(h\omega)  - 1)^2  - \frac{1}{n} \mcK(h\omega)^2  - 1\right ) \mcF(\omega)^2 d\omega + \frac{1}{nh}\int  \mcK(\omega)^2 \ d\omega \right ] \\
& = & \left ( 1 - \frac{1}{n} \right ) \left ( \MISE_n(h)  - \int \mcF(\omega)^2  \ d\omega \right )\\
& = & \left ( 1 - \frac{1}{n} \right ) \left ( \MISE_n(h)  - \int f(x)^2  \ dx \right ).
\end{eqnarray*}
\el
This implies that, modulo a shifting and scaling factor that are both independent of $h$, the expectation of our estimator is exactly $\MISE_n(h)$.  Additionally, it becomes clear that this function must attain its minimum at the same value of $h$ as $\MISE_n(h)$, and modulo a shift we have
$ \E \left [\mcE_n(h) \right ] \sim \MISE_n(h)$ as $n \to \infty$.

\section{Appendix C: Numerical Bandwidth Estimation}\label{sec:numerics}

Next, we outline a numerical approach based on our use of the Fourier transform. Implementing an iterative algorithm to compute the approximate distribution, let us assume that the algorithm converges, and hence the final density and the kernel must converge to the same function, while the bandwidth must also converge to some $h > 0$. Then, denoting the estimated kernel (based on data) by $K(x)$, this function must satisfy
\bl
\begin{equation*}
K(x) =\frac1n \sum_{j=1}^n\frac1h K \biggl( \frac{x-X_j}{h} \biggr).
\end{equation*}
\el
Additionally, its Fourier transform then satisfies the relationship
\bl
\begin{align*}
\hat K(\omega) &= \frac1{nh} \int e^{-2\pi i \omega x} \sum_{j=1}^n K \left( \frac{x-X_j}{h} \right) dx \\
&=\frac1{nh} \int \sum_{j=1}^n e^{-2\pi i \omega(zh+X_j)} K (z) h dz \\
&= \frac1n  \sum_{j=1}^n e^{-2\pi i \omega X_j} \int e^{-2\pi i \omega zh}K (z) dz \\
&= \frac1n \sum_{j=1}^n e^{-2\pi i \omega X_j} \hat K (h\omega) \\
&= \hat{f}_n(\omega )\hat K(h\omega )
\end{align*}
\el
due to \eqref{eq:FTf}.
With this and the asymptotic approximation \eqref{EFn}, we have
\bl
\beq
\label{eq:Khat}
\hat K(\omega) = \hat{f}(\omega )\hat K(h\omega ) + \mathcal{O}\left ( \frac{1}{\sqrt{n}} \right )
\eeq
\el
for $n$ suitably large.
Revisiting the expression for the MISE \eqref{eq:FTMISE} yields
\bl
\begin{equation*}
\MISE_n(h) =  \int \left (\mcK(h\omega)-1 \right)^2 \mcF(\omega)^2 \ d\omega + \mathcal{O}\left ( \frac{1}{n} \right )
\end{equation*}
\el
for $n$ suitably large.
Expanding this expression and using the relationship \eqref{eq:Khat} satisfied by the Fourier transforms of the limiting kernel and unknown pdf, we find
\bl
\begin{align*}
\MISE_n(h) 
& =  \int \left (\mcK(h\omega)^2\mcF(\omega)^2  - 2 \mcK(h\omega)\mcF(\omega)^2 + \mcF(\omega)^2\right) \ d\omega + \mathcal{O}\left ( \frac{1}{n} \right )\\
& = \int \left (\mcK(\omega)^2  - 2 \mcK(\omega)\mcF(\omega) + \mcF(\omega)^2\right) \ d\omega + \mathcal{O}\left ( \frac{1}{\sqrt{n}} \right )\\ 
& = \int \left | \mcK(\omega)  - \mcF(\omega)\right |^2 d\omega + \mathcal{O}\left ( \frac{1}{\sqrt{n}} \right ).
\end{align*}
\el
Therefore, we see that the iterative algorithm guarantees that the $\MISE$ is minimized precisely when the kernel $K(x)$ converges to the unknown distribution in the $L^2$ sense.
This implies that when the algorithm converges it must converge to the unknown pdf $f(x)$ because an unbiased estimator for the $\MISE$ is minimized at every step.

\begin{figure}
 \centering
 \includegraphics[width=5.5in]{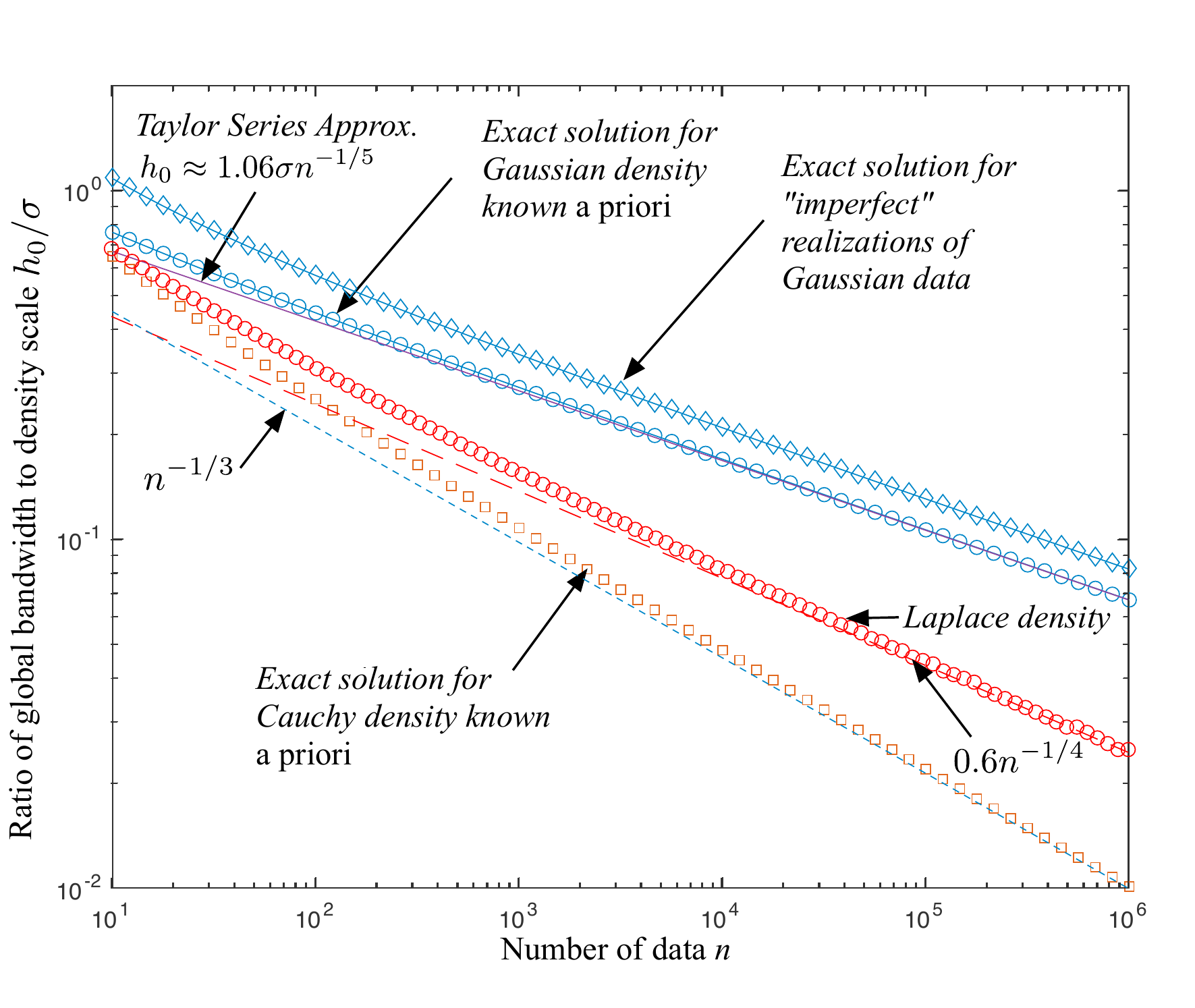}
 \caption{Log-log plots of global bandwidth over data scale parameter $(h_0/\sigma)$ versus number $n$ of data using exact solution for Gaussian data density known {\em a priori} (Eq. \eqref{eq:eps_Gauss}) versus Taylor series approximate solution (Eq. \eqref{eq:Taylor}) and numerical estimation of Gaussian data density (Eq. \eqref{eq:epsdef}). Also shown are the lower values of $h_0/\sigma$ for Cauchy data estimated with Cauchy kernel. }
 \label{fig:h_0_exact}
 \end{figure}

Furthermore, our analysis now demonstrates the appropriate range of Taylor series estimates of $h$, because the exact result can be derived from the Fourier transform.  
Assume a Gaussian for the kernel and also assume {\em a priori} that the underlying data are Gaussian (with zero mean and variance $\sigma^2$), so that $\hat K(\omega)=\exp(-(2\pi\omega)^2/2)$ and $\hat f(\omega) = \exp(-\sigma^2(2\pi\omega)^2/2)$.  The first integral in Eq. \eqref{eq:FTMISE} can be computed in several ways, but is easily performed by recognizing the form of a Gaussian, so that
\bl
\begin{equation*}
\frac{1}{nh}\int\ e^{-(2\pi\omega)^2} d\omega=\frac1{nh} \frac{1}{\sqrt{4\pi} }\int \frac{1}{\sqrt{2\pi/(8\pi^2)}} e^{\left( \frac{- \omega^2} {2/(8\pi^2)} \right) }  d\omega=\frac{1}{2\sqrt{\pi}nh},
\end{equation*}
\el
owing to the fact that the last integral is of a density in $\omega$.  Similarly, the second integral in Eq. \eqref{eq:FTMISE} is 
\bl
\begin{align*}
&\int \left ( (1-1/n) \hat K^2(h\omega) - 2\hat K(h\omega) \right ) \hat f^2(\omega) d\omega\\
&= \int \left ((1-1/n)e^{-(2\pi h \omega)^2}-2e^{-(2\pi h \omega)^2/2} \right )e^{-\sigma^2(2\pi \omega)^2}d\omega \\
&= \int \left ((1-1/n)e^{-(\sigma^2+h^2)(2\pi \omega)^2}-2e^{-(\sigma^2+h^2/2)(2\pi \omega)^2} \right )d\omega\\
& = \frac{(1-1/n)}{2\sqrt{\pi(h^2+\sigma^2)}} - \frac{1}{\sqrt{\pi(\sigma^2+h^2/2)}}
\end{align*}
\el
where we have rearranged as before to make Gaussian densities (in $\omega$) for each term.  Therefore, the {resulting} quantity to be minimized is now 
\bl
\beq\label{eq:eps_Gauss}
\MISE_n(h)=\frac{1}{2\sqrt{\pi}nh}+\frac{1}{2\sqrt{\pi(\sigma^2+h^2)}}-\frac{1}{2n\sqrt{\pi(\sigma^2+h^2)}}-\frac{1}{\sqrt{\pi( \sigma^2+h^2/2)} }.
\eeq
\el
It suffices to approximate the $h_0$ that minimizes $\MISE_n(h)$ to any numerical tolerance, by taking $d(\MISE_n(h))/dh$, setting {it} to zero, and finding the root of the resulting equation. As expected, the estimate of $h_0$ based on Taylor series is worse for smaller data sets (i.e., $n \lesssim 100$), but as $n$ grows large, the Taylor series solution converges to the exact solution (Fig. \ref{fig:h_0_exact}).   
However, it is important to note that these quantities are the optimal bandwidth when both the kernel {\em and the underlying data density} are known to be Gaussian.  If the underlying density is unknown, then the data are used to construct the quantity to be minimized $\varepsilon_n(h)$ in Eq. \eqref{eq:epsdef}.  To see how this differs, we can imagine that perfectly Gaussian data is generated.  Then Eq. \eqref{eq:epsdef} evaluates to
\bl
\beq
\label{eq:eps_Gauss}
\varepsilon_n(h)=\frac{2}{\sqrt{2\pi}nh}+\frac{1}{2\sqrt{\pi(\sigma^2+h^2)}}-\frac{1}{2n\sqrt{\pi(\sigma^2+h^2)}}-\frac{1}{\sqrt{\pi( \sigma^2+h^2/2)} },
\eeq
\el
which has a root approximately $4^{1/5}=1.32$ larger for large $n$ (Fig. \ref{fig:h_0_exact}).  The fact that data are imperfect means that the global bandwidth must be about 32\% to 70\% larger (depending on $n$) to achieve additional smoothing when compared to a completely ``perfect'' realization of data.  

\begin{figure}
 \centering
 \includegraphics[width=5.5in]{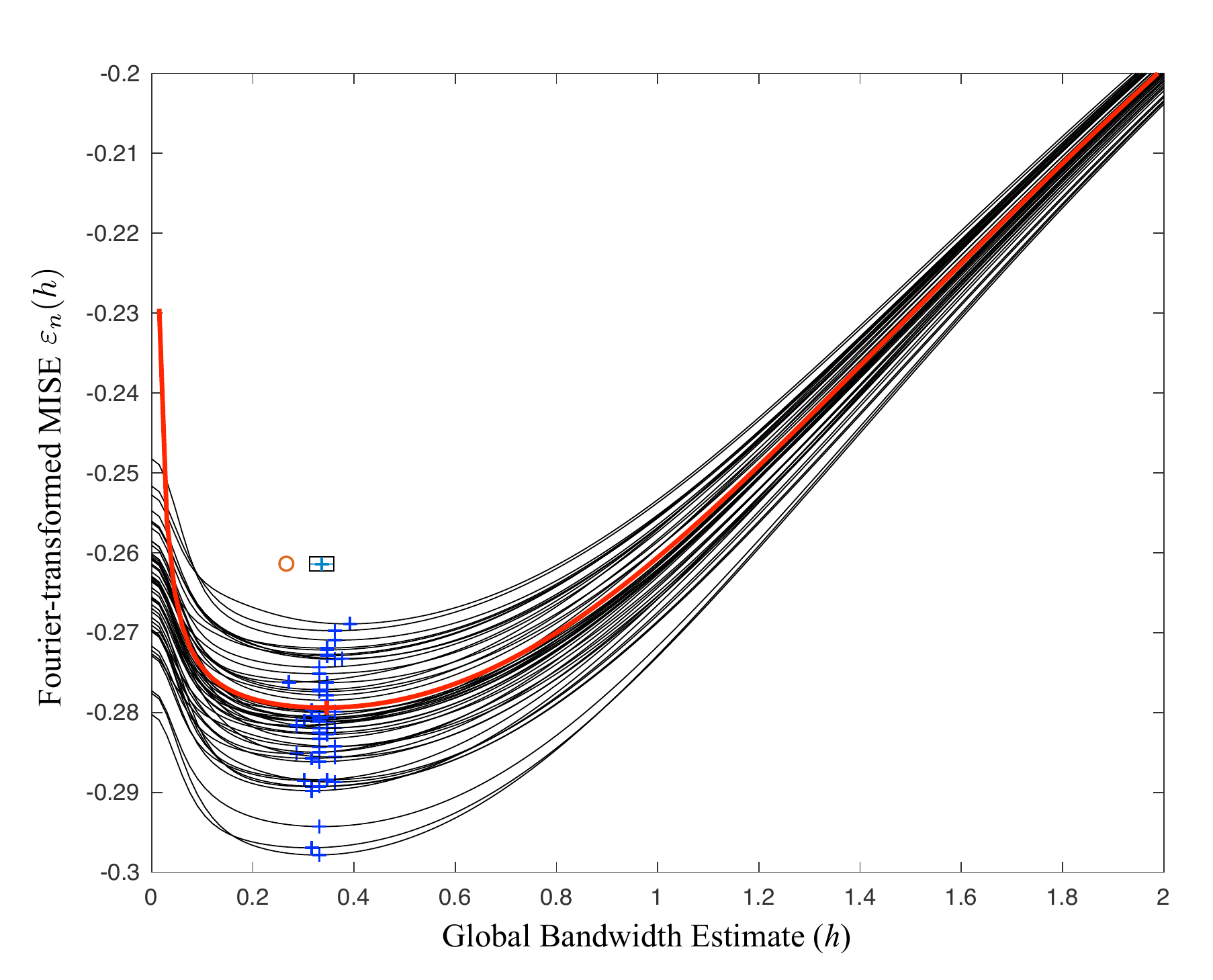}
 \caption{Plots of unbiased estimator for Fourier-transformed MISE (denoted $\varepsilon_n(h)$) as a function of  global bandwidth parameter $h$. Plots either assume or use Gaussian data with $\sigma^2=1$ and 1,000 data points.  The exact expression Eq. \eqref{eq:eps_Gauss} is plotted with a thick red curve; the minimum (shown with a $+$ sign) is found at $h_0=0.3406$.   The estimate of $h_0$ using Taylor series is 0.266 and is denoted by a circle above the curves.  Also shown is an ensemble of 50 curves (in black) wherein for each curve 1,000 IID Gaussian data are generated and the density function is estimated by Fourier transform Eq. \eqref{eq:FTf}.  The ensemble statistics of the estimated $h_0$ were calculated with mean $h_0=0.335$, with standard deviation of $\sigma_{h_0}=0.0238$ (box above  curves denotes mean $\pm\sigma_{h_0}$). }
 \label{fig:eps_Gauss}
 \end{figure}

For a specific example, we set $n=1000$ and $\sigma^2=1$, which produces an exact optimal global bandwidth (for imperfect data using $\varepsilon_n(h)$ in Eq. \eqref{eq:epsdef}) of $h_0=0.341$ (Figure \ref{fig:eps_Gauss}), whereas the estimate based on Taylor expansions gives $h_0 = 1.06\sigma n^{-1/5} = 0.266$.  It is important to see how well a numerical estimate of the data density gives an estimate of $h_0$, rather than simply assuming a Gaussian density function.  We may now compare the values of $h$ that are estimated using the Fourier-transformed data to form an estimate of the density function (i.e., using Eq. \eqref{eq:FTf} in Eq. \eqref{eq:epsdef}) instead of assuming the Gaussian form.  Here we show the results for 50 independent runs in which 1,000 IID Gaussian data are generated and the experimental curve generated and $h$ taken at the curve minimum (black curves in Figure \ref{fig:eps_Gauss}).  While there is a large vertical spread in the curves, the locations of the minima are fairly tightly constrained.  The mean of 50 values of $h_0$ is 0.335 (compared to the exact value of 0.341), and the estimated $h_0$ have a standard deviation of 0.0238.

Several other characteristic functions (Fourier transforms of PDFs) are easily integrated and illustrate the effect of data distribution on estimation of $h_0$.  For example, the standard Cauchy density, defined by
\bl
\begin{equation*}
f(x) = \frac{\sigma}{\pi \left ( \sigma^2 + x^2 \right)},
\end{equation*}
\el
has both divergent variance and mean, and its associated Fourier transform is given by 
\bl
\begin{equation*}
\hat f(\omega) = \exp(-2\pi\sigma|\omega|).
\end{equation*}
\el
Note that the scale parameter $\sigma$ commonly used for stable densities is not the standard deviation, which is infinite.  Assuming that the kernel was also a perfect copy of the data density, so that 
\bl
\begin{equation*}
\hat K(\omega)=\exp(-2\pi|\omega|),
\end{equation*}
\el
{the MISE becomes (up to an additive factor independent of $h$)
\bl
\begin{equation*}
\text{MISE}_n(h) = \frac{1}{2\pi nh} + \left (1 - \frac{1}{n} \right ) \frac{1}{2\pi (h + \sigma)} - \frac{1}{\pi  \left (\frac{1}{2}h + \sigma \right )}.
\end{equation*}
\el
Therefore,} calculating the minimum of the MISE \eqref{eq:FTMISE} means solving the root of 
\bl
\beq\label{eq:h_Cauchy}
\frac{d (\text{MISE}_n(h))}{dh} = -\frac{1}{2\pi n} - \left(1-\frac{1}{n} \right)\frac{1}{2\pi \left (1+\frac{\sigma}{h}\right )^2} + \frac{2}{\pi \left (1+\frac{2\sigma}{h} \right)^2}.
\eeq
\el
These values of $h_0(n)$ are significantly smaller than those found for Gaussian data (Fig. \ref{fig:h_0_exact}) and also decline for large $n$ approximately like $\sim n^{-1/3}$.  This suggests a numerical procedure for simultaneous estimation of the data density and the global bandwidth.  The FT estimate of $h_0$ based on Gaussian data is the largest of the estimates (Fig. \ref{fig:h_0_exact}), so we begin with that value.  If the iterated kernel---based on the estimated density and using this $h_0$---fails to converge, then we reduce $h_0$ systematically down to a minimum given by the Cauchy $h_0$.  In this procedure, the specifics of the data distribution need not be known.  Simply start with an assumption of Gaussian-like smoothness and data density, but allow for Cauchy-like sparcity of data (i.e., few very large data).  

We may also consider the Laplace (or double exponential) density defined by
\bl
\begin{equation*}
f(x) = \frac{1}{2\sigma} \exp\left (-\frac{|x|}{\sigma} \right ), 
\end{equation*}
\el
which has mean zero and variance $2\sigma^2$, but does not possess a continuous derivative at $x = 0$.
The Fourier transform of this function is given by 
\bl
\begin{equation*}
\hat f(\omega) = \frac{1}{1 + 4\pi^2 \sigma^2 \omega^2}.
\end{equation*}
\el
If the kernel is similarly distributed so that 
\bl
\begin{equation*}
\hat K(\omega)=\frac{1}{1 + 4\pi^2\omega^2},
\end{equation*}
\el
then the MISE becomes
\bl
\begin{equation*}
\text{MISE}_n(h) = \frac{1}{4nh} + \left (1 - \frac{1}{n} \right ) \frac{h^2 + 3h\sigma + \sigma^2}{4(h + \sigma)^3} - \frac{2h + \sigma}{2(h + \sigma)^2}.
\end{equation*}
\el
As before, the minimum of the MISE \eqref{eq:FTMISE} can be computed by finding the root of the derivative of this expression, namely
\bl
\beq\label{eq:h_Laplace}
\frac{d (\text{MISE}_n(h))}{dh} = -\frac{1}{4n}\left ( 1 + \frac{\sigma}{h} \right )^2 + \left ( \frac{3}{4} + \frac{1}{4n}  \right ) \frac{\frac{h}{\sigma}}{\frac{h}{\sigma}+1} - \left ( \frac{3}{4} - \frac{1}{4n}  \right ) \frac{\frac{h}{\sigma}}{\left (\frac{h}{\sigma}+1 \right )^2}.
\eeq
\el
The resulting values of $h_0(n)$ are again significantly smaller than those found for Gaussian data (Fig. \ref{fig:h_0_exact}) and also decline for large $n$ approximately like $\sim n^{-1/4}$.  
%

\begin{figure}
 \centering
 \includegraphics[width=5.5in]{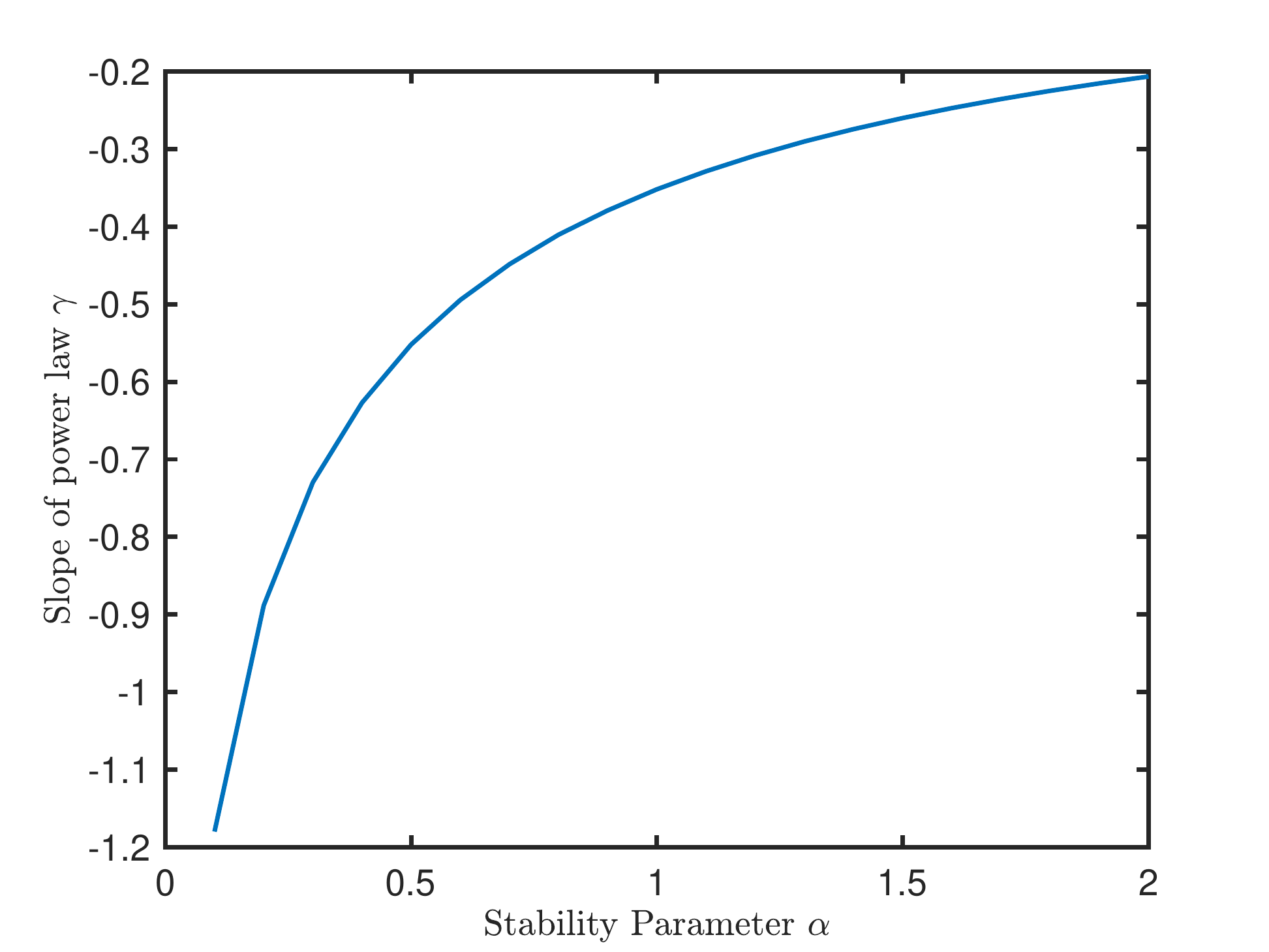}
 \caption{ Slope of power law decline $\sim n^{-\gamma}$ of optimal kernel bandwidth with particle number for stable distributions with different stability parameter $\alpha$}
 \label{fig:stables}
 \end{figure}

Finally, we consider the family of stable distributions, whose density may be defined by 
\bl
\begin{equation*}
f(x) = \frac{1}{2\pi} \int_{-\infty}^{\infty} e^{-\vert c k\vert^\alpha (1-i\beta sgn(k) \tan(\frac{\pi \alpha}{2}))} e^{-ikx}dk 
\end{equation*}
\el
where $0<\alpha\leq 2$ is the stability parameter, $-1<\beta<1$ is a skewness parameter and $c$ is the scale parameter. The Fourier transform is given by 
\bl
\begin{equation*}
\hat f(\omega)=e^{-\vert 2\pi c \omega\vert^\alpha (1-i\beta sgn(-\omega) \tan(\frac{\pi \alpha}{2}))}.
\end{equation*}
\el
As before, if the kernel is similarly distributed so that 
\bl
\begin{equation*}
\hat K(\omega)=e^{-\vert 2\pi \omega\vert^\alpha (1-i\beta sgn(-\omega) \tan(\frac{\pi \alpha}{2})}
\end{equation*}
\el
then the MISE becomes
\bl
\begin{equation*}
\text{MISE}_n(h) = C \bigg[ \frac{2^{-1/\alpha}}{nh}+\bigg(1-\frac{1}{n}\bigg) \bigg(2h^\alpha+2c^\alpha\bigg)^{-1/\alpha}-2\bigg(h^\alpha+2c^\alpha\bigg)^{-1/\alpha}\bigg] 
\end{equation*}
\el
where
\bl
\begin{equation*}
C=\int_{-\infty}^\infty e^{-\vert 2\pi \omega\vert^\alpha (1-i\beta sgn(-\omega) \tan(\frac{\pi \alpha}{2}))} d\omega
\end{equation*}
\el
As before this can be minimized and, similar to the distributions explored so far, we find that there is a power law decline $\sim n^{-\gamma}$ where $\gamma$ depends on $\alpha$ as depicted in Figure \ref{fig:stables}. Note that the magnitude of $\MISE$ depends on $\alpha$ and $\beta$ through the constant $C$, but that this does not impact the minimized value.
We also note that these calculations may be made for other densities but are not shown.  Additionally, some of the integrations must be performed numerically as it may be the case that no closed-form expression for the antiderivative exists.  

\newpage
REFERENCES
\bibliography{reaction_proposal}

\end{document}